\documentclass[a4paper,fleqn,usenatbib]{mnras}

\usepackage{etoolbox}
\makeatletter
\newcount\c@additionalboxlevel
\newcount\c@maxboxlevel
\patchcmd\@combinedblfloats{\box\@outputbox}{%
  \stepcounter{additionalboxlevel}%
  \box\@outputbox
}{}{\errmessage{\noexpand\@combinedblfloats could not be patched}}

\AtBeginShipout{%
  \ifnum\value{additionalboxlevel}>\value{maxboxlevel}%
    \typeout{Warning: maxboxlevel might be too small, increase to %
      \the\value{additionalboxlevel}%
    }%
  \fi 
  \@whilenum\value{additionalboxlevel}<\value{maxboxlevel}\do{%
    \typeout{* Additional boxing of page `\thepage'}%
    \setbox\AtBeginShipoutBox=\hbox{\copy\AtBeginShipoutBox}%
    \stepcounter{additionalboxlevel}%
  }%
  \setcounter{additionalboxlevel}{0}%
}
\makeatother

\usepackage[T1]{fontenc}
\usepackage{ae,aecompl}

\usepackage{graphicx,rotating,pdflscape,geometry}	
\usepackage{amsmath}	
\usepackage{amssymb}	

\title[The ISM in Andromeda's dSphs: II. Multi-phase gas content]{The interstellar medium in Andromeda's dwarf spheroidal galaxies: II. Multi-phase gas content and ISM conditions} 
\author[De Looze et al.]{Ilse De Looze$^{1,2,3}$\thanks{E-mail: idelooze@star.ucl.ac.uk, ilse.delooze@ugent.be}, Maarten Baes$^{2}$ , Diane Cormier$^{4}$, Hiroyuki Kaneko$^{5}$, Nario Kuno$^{6}$,
\newauthor Lisa Young$^{7,8}$ George J. Bendo$^{9}$, M{\'e}d{\'e}ric Boquien$^{3,10}$,  Jacopo Fritz$^{2,11}$, Gianfranco 
\newauthor Gentile$^{2}$,  Robert C. Kennicutt$^{3}$, Suzanne C. Madden$^{12}$,  Matthew W.~L. Smith$^{13}$
\newauthor and Christine D. Wilson$^{14}$ \\
$^{1}$ Department of Physics and Astronomy, University College London, Gower Street, London WC1E 6BT, UK \\
$^{2}$ Sterrenkundig Observatorium, Universiteit Gent, Krijgslaan 281 S9, B-9000 Gent, Belgium \\
$^{3}$ Institute of Astronomy, University of Cambridge, Madingley Road, Cambridge, CB3 0HA, UK \\
$^{4}$ Zentrum f{\"u}r Astronomie der Universit{\"a}t Heidelberg, ITA, Albert-Ueberle Str. 2, D-69120 Heidelberg, Germany \\ 
$^{5}$ Nobeyama Radio Observatory, Minamimaki, Minamisaku, Nagano 384-1305 \\
$^{6}$ Faculty of Pure and Applied Sciences, University of Tsukuba, 1-1-1 Tennoudai, Tsukuba, Ibaraki 350-8577, Japan \\
$^{7}$ Physics Department, New Mexico Institute of Mining and Technology, Socorro, NM 87801, USA \\
$^{8}$ Academia Sinica Institute of Astronomy \& Astrophysics, PO Box 23-141, Taipei 10617, Taiwan, R.O.C. \\
$^{9}$ UK ALMA Regional Centre Node, Jodrell Bank Centre for Astrophysics, School of Physics and Astronomy, \\ University of Manchester, Oxford Road, Manchester M13 9PL, United Kingdom \\
$^{10}$ Unidad de Astronom{\'i}a, Fac. Cs. B{\'a}sicas, Universidad de Antofagasta, Avda. U. de Antofagasta 02800, Antofagasta \\
$^{11}$ Centro de Radioastronom{\'i}a y Astrof{\'i}sica, CRyA, UNAM, Campus Morelia, A.P. 3-72, C.P. 58089 Michoac{\'a}n, Mexico \\
$^{12}$ Laboratoire AIM, CEA, Universit{\'e} Paris VII, IRFU/Service d$'$Astrophysique, Bat. 709, 91191 Gif-sur-Yvette, France \\
$^{13}$ School of Physics and Astronomy, Cardiff University, Queens Buildings, The Parade, Cardiff CF24 3A A \\
$^{14}$ Department of Physics \& Astronomy, McMaster University, Hamilton, ON L8S 4M1 Canada 
}

\date{Accepted 2016 November 16. Received 2016 November 15; in original form 2016 September 27.}

\pubyear{2016}

\begin{document}
\label{firstpage}
\pagerange{\pageref{firstpage}--\pageref{lastpage}}
\maketitle

\begin{abstract}
We make an inventory of the interstellar medium material in three low-metallicity dwarf spheroidal galaxies of the Local Group (NGC\,147, NGC\,185 and NGC\,205). Ancillary H~{\sc{i}}, CO, \textit{Spitzer} IRS spectra, H$\alpha$ and X-ray observations are combined to trace the atomic, cold and warm molecular, ionised and hot gas phases. We present new Nobeyama CO(1-0) observations and \textit{Herschel} SPIRE FTS [C~{\sc{i}}] observations of NGC\,205 to revise its molecular gas content.  

We derive total gas masses of $M_{\text{g}}$ = 1.9-5.5$\times$10$^{5}$ M$_{\odot}$ for NGC\,185 and $M_{\text{g}}$ = 8.6-25.0$\times$10$^{5}$ M$_{\odot}$ for NGC\,205. Non-detections combine to an upper limit on the gas mass of $M_{\text{g}}$ $\leq$  0.3-2.2$\times$10$^{5}$ M$_{\odot}$ for NGC\,147. The observed gas reservoirs are significantly lower compared to the expected gas masses based on a simple closed-box model that accounts for the gas mass returned by planetary nebulae and supernovae. The gas-to-dust mass ratios GDR$\sim$37-107 and GDR$\sim$48-139 are also considerably lower compared to the expected GDR$\sim$370 and GDR$\sim$520 for the low metal abundances in NGC\,185 (0.36\,Z$_{\odot}$) and NGC\,205 (0.25\,Z$_{\odot}$), respectively.

To simultaneously account for the gas deficiency and low gas-to-dust ratios, we require an efficient removal of a large gas fraction and a longer dust survival time ($\sim$1.6\,Gyr). We believe that efficient galactic winds (combined with heating of gas to sufficiently high temperatures in order for it to escape from the galaxy) and/or environmental interactions with neighbouring galaxies are responsible for the gas removal from NGC\,147, NGC\,185 and NGC\,205. 
\end{abstract}



\section{Introduction}
Dwarf spheroidal galaxies (dSph) dominate the overall galaxy population in the Universe at the low mass end. With most dwarf spheroidals residing in groups and clusters of galaxies, environmental effects are thought to play an important role in the formation and evolution of the dSph galaxy population. Studying the properties of the interstellar medium in dwarf spheroidal galaxies, in combination with their star formation histories, can give us clues to their formation processes (e.g., \citealt{2009ARA&A..47..371T}) and the role of environmental processes in their evolution (e.g., \citealt{2008ApJ...674..742B}). Being the most prominent dSph residents of the Local Group, the three dwarf satellites of Andromeda, NGC\,147, NGC\,185 and NGC\,205, offer the best opportunity to study the interstellar medium of dwarf spheroidal galaxies in the nearby Universe.  
 
In \citet{2016MNRAS.459.3900D}, we focused on the dust reservoirs in the three dSph satellites of Andromeda and show that the observed dust masses in NGC\,185 and NGC\,205 are significantly higher compared to the estimated metal enrichment from evolved stars and supernova remnants. Although uncertainties on the dust yields from asymptotic giant branch (AGB) and supernovae might affect the estimated dust production rates, the observed dust masses exceed predictions by an order of magnitude and can only be explained by efficient interstellar grain growth or longer dust survival times (3-6\,Gyr).

Based on observational constraints of the evolved stellar populations, the dSph satellites NGC\,147, NGC\,185 and NGC\,205 are shown to be characterised by significantly lower gas masses compared to the predicted material returned by evolved stars and the left-over gas reservoir that remains after previous star-formation episodes \citep{1998ApJ...507..726S,1998ApJ...499..209W,2012MNRAS.423.2359D}. The gas deficiency in NGC\,205 was attributed to environmental interactions with parent galaxy Andromeda and/or efficient stellar feedback \citep{2012MNRAS.423.2359D}. To rule out that an important gaseous ISM reservoir has been overlooked in previous studies, we require an accurate quantification of the interstellar material (its mass and properties) in dSphs in combination with models that account for the gas mass returned by the evolved stellar population and supernovae. In this paper, we make a revised inventory and updated analysis of the gaseous reservoir in the three dSph satellite galaxies of Andromeda by taking into account all significant phases of their ISM. We present a new Nobeyama CO(1-0) map of the southern regions in NGC\,205, \textit{Herschel} PACS line spectroscopy observations for NGC\,185 and \textit{Herschel} SPIRE FTS spectroscopy observations for NGC\,205. We furthermore use ancillary data of other gaseous components tracing the atomic gas (H~{\sc{i}}), cold (CO), CO-dark ([C~{\sc{i}}]) and warm molecular gas (H$_{2}$ rotational lines), ionised gas (H$\alpha$), and hot X-ray emitting gas. 

The description of the star formation histories and characterisation of the most recent star formation rates and metal abundances for the three galaxies under study have been outlined in the three paragraphs below. Table \ref{GeneralProp} presents an overview of the general properties and available observational constraints for each of the galaxies. In Section \ref{Data.sec}, we present the new NRO 45m CO(1-0) observations, \textit{Herschel} PACS and SPIRE spectroscopy data, and the ancillary datasets used to analyse the gaseous reservoirs in NGC\,147, NGC\,185 and NGC\,205. To learn more about the physical gas conditions, we analyse the origin of the [C~{\sc{ii}}] emission in NGC\,185 (\ref{Origin.sec}), quantify the photoelectric efficiency (\ref{PE.sec}), and compare the emission of gas tracers to photo-dissociation models (\ref{PDR.sec}). Several observations are combined to derive the total gas content in the three dwarf spheroidal galaxies (Section \ref{Totalgas.sec}). Section \ref{SFE.sec} investigates the position of dSphs on the local Kennicutt-Schmidt relation. Combining dust and gas mass reservoirs, we derive gas-to-dust mass ratios for NGC\,185 and NGC\,205 in Section \ref{GDR.sec}. The ISM mass budget in the three dwarf spheroidal companions of Andromeda (NGC\,147, NGC\,185, NGC\,205) is compared to a simple closed box model and discussed in light of galaxy evolution processes in Section \ref{MissingMass.sec}. The main results are summarised in Section \ref{Conclusions.sec}. 
Throughout this paper, we adopt distances of $675\pm27$ kpc, $616\pm26$ kpc and $824\pm27$ kpc to NGC\,147, NGC\,185, and NGC\,205 \citep{2005MNRAS.356..979M}, respectively. 

\subsection{Star formation history}
\label{SFH.sec}
Dwarf spheroidal galaxies are considered to form their stars in a limited number of star formation episodes lasting a few Gyr and clearly separated by quiescent periods (e.g., \citealt{2004MNRAS.351.1338L}). \citet{2012MNRAS.419.3159M} determined that the star formation in NGC\,185 has taken place in three major episodes separated by quiescent periods without any significant star formation activity. The first SF episode, during which most of the stellar content was produced, took place $\sim$ 10 Gyr ago in NGC\,185 \citep{2015ApJ...811..114G}, resulting in a stellar population with an iron abundance of [Fe/H] $\sim$ -1.0. After the first star formation episode, which lasted a few Gyr, NGC\,185 had a long quiescent period without any significant star formation activity. The presence of an intermediate age population (2-3 Gyr old) suggested a secondary star formation episode. This second cycle of star formation was considered to be the result of the build-up of mass loss from evolved stars and/or planetary nebulae \citep{1996ApJ...470..781W,2005AJ....130.2087D}. A similar old and intermediate stellar population has been observed in NGC\,147. The old stellar population in NGC\,147 has a mean age (7.5 Gyr) and metallicity ([Fe/H] $\sim$ -0.7), making the stars considerably younger in this galaxy and more metal-rich compared to the stars in NGC\,185 (with mean age of 10 Gyr and [Fe/H] $\sim$ -1.0). This suggests that the bulk of stars in NGC\,185 already formed at an earlier epoch \citep{2015ApJ...811..114G}. In the central regions of NGC\,185, a more recent star formation episode took place that started a few 100\,Myr ago. NGC\,147 shows no signs of any recent star formation activity \citep{1997AJ....113.1001H}.  An old stellar population (10 Gyr, \citealt{1990A&A...228...23B}) also dominates the overall stellar content of NGC\,205, while a plume of bright blue star clusters in the central region of this galaxy was already identified 60 years ago \citep{1951POMic..10....7B,1973ApJ...182..671H}.

\subsection{Star formation rates}
\label{SFR.sec}
In NGC\,147, no significant star formation activity has taken place during the last 1 Gyr \citep{1997AJ....113.1001H}. The star formation rate (SFR $\sim$ 6.6 $\times$ 10$^{-4}$ M$_{\odot}$ yr$^{-1}$) in the central regions of NGC\,185 (inner 118$\arcsec$) over the last $\sim$ 1 Gyr has been determined from color-magnitude diagrams by \citet{1999AJ....118.2229M}. The total SFR $\sim$ 82 $\times$ 10$^{-4}$ M$_{\odot}$ yr$^{-1}$ (over the entire lifetime of the galaxy) in those central regions is significantly higher and consistent with a star formation history where most of the stars have been formed in the first few Gyr after the formation of the galaxy \citep{1999AJ....118.2229M}. The latter central SFR should also be considered as an upper limit given that the inner regions are affected by crowding and every blue object has been assumed to be an individual star. In a similar way, the SFR ($\sim$ 7.0 $\times$ 10$^{-4}$ M$_{\odot}$ yr$^{-1}$) for the central 28$\arcsec$ $\times$ 26$\arcsec$ region in NGC\,205 has been derived from color-magnitude diagrams for the stars produced between $\sim$ 62 and $\sim$ 335 Myr ago over a time period of $\sim$ 273 Myr \citep{2009A&A...502L...9M}. The latter SFR estimates derived from colour-magnitude diagrams should be more reliable compared to standard SFR calibration, which require sufficient sampling of stellar ages and assume a continuous star formation activity.

\subsection{Metal abundance determination}
\label{Metal.sec}
For NGC\,147, the mean metallicity 12+$\log$(O/H) $\sim$ 8.06 (or 0.23 Z$_{\odot}$) is estimated from observations of eight planetary nebulae \citep{2007MNRAS.375..715G}. The latter metallicity based on planetary nebulae (PNe) is not very different from the metallicity derived for the old stellar population ([Fe/H] $\sim$ -0.7, \citealt{2015ApJ...811..114G}), which is in line with a low star formation activity and negligible metal enrichment during the last few Gyr in NGC\,147.
The metallicity in NGC\,185 is estimated by averaging the oxygen abundances derived for 5 central PNe reported by \citet{2008ApJ...684.1190R}, resulting in 12+$\log$(O/H) $\sim$ 8.25 or Z $\sim$ 0.36 $Z_{\odot}$ (assuming a solar oxygen abundance of 12+$\log$(O/H) $\sim$ 8.69, \citealt{2009ARA&A..47..481A}). Similarly, \citet{2012MNRAS.419..854G} find a mean oxygen abundance of 12+$\log$(O/H) $\sim$ 8.20 or Z $\sim$ 0.32 $Z_{\odot}$ for NGC\,185 based on independent observations for four of the same planetary nebulae. 
In the same way, we derive the mean oxygen abundance 12+$\log$(O/H) $\sim$ 8.08 (or 0.25 Z$_{\odot}$\footnote{The latter mean oxygen abundance is a bit higher compared to the mean value (12+$\log$(O/H) $\sim$ 7.80) used in \citet{2012MNRAS.423.2359D} which was calculated as the average of thirteen planetary nebulae analysed by \citet{2008ApJ...684.1190R}.} based on fourteen planetary nebulae in NGC\,205 \citep{2014MNRAS.444.1705G}.

We caution that the abundances of planetary nebulae (probing the evolutionary products of the intermediate mass stars) might be lower with respect to the abundances in H~{\sc{ii}} regions (which probe the initial phases of massive stellar evolution) due to their different stages of evolution. Based on comparison studies of elemental abundances derived for H~{\sc{ii}} regions and PNe in NGC\,300 \citep{2013A&A...552A..12S} and M\,33 \citep{2010MNRAS.404.1679B,2010A&A...512A..63M}, we consider a maximum offset of 0.15 dex between the abundances from H~{\sc{ii}} regions and PNe. 

\begin{table}
\caption{Overview of the galaxy characteristics and availability of observational constraints for the three dwarf spheroidal galaxies NGC\,147, NGC\,185 and NGC\,205. Total galaxy masses (including baryonic and dark matter) have been taken from \citet{2006MNRAS.369.1321D}. The star formation rates are representative for the star formation activity during the last 1\,Gyr (see Section \ref{SFR.sec}). The metallicity is constrained by the average oxygen abundance of planetary nebulae (see Section \ref{Metal.sec}).
The symbols indicate that the galaxy has been observed and detected ("x"), observed but not detected ("$<$") or was not observed ("o") for that specific gas tracer.}
\label{GeneralProp}
\centering
\begin{tabular}{|l|ccc|}
\hline 
Galaxy properties & NGC\,147 & NGC\,185 & NGC\,205 \\
 \hline
Distance [kpc] & 675 & 616 & 824 \\
Galaxy mass [10$^{8}$ M$_{\odot}$] & 3.0 & 2.6 & 10.2 \\
SFR [M$_{\odot}$ yr$^{-1}$] & 0* & 6.6$\times$10$^{-4}$ & 7.0$\times$10$^{-4}$ \\
Metallicity [12+$\log$O/H] & 8.06 & 8.25 & 8.08 \\
\hline 
Observational constraints & NGC\,147 & NGC\,185 & NGC\,205 \\
 \hline
H{\sc{i}} & < & x & x \\
CO (1-0) & < & x & x \\
warm H$_{2}$ (\textit{Spitzer} IRS) & o & x & o \\
$[$C{\sc{ii}}], [O{\sc{i}}] (\textit{Herschel} PACS) & o & x & x \\ 
$[$C{\sc{i}}] (\textit{Herschel} SPIRE FTS) & o & o & $<$ \\
$[$N{\sc{ii}}] (\textit{Herschel} SPIRE FTS) & o & o & $<$ \\
H$\alpha$ & o & x & $<$  \\
X-ray & $<$ & $<$ & $<$  \\
 \hline \\
\end{tabular}	
*There is not evidence of any recent star formation activity in NGC\,147 \citep{1997AJ....113.1001H}.
\end{table}

\section{Data}
\label{Data.sec}

\subsection{PACS spectroscopy data of NGC\,185}
\begin{figure}
\centering
\includegraphics[width=8.5cm]{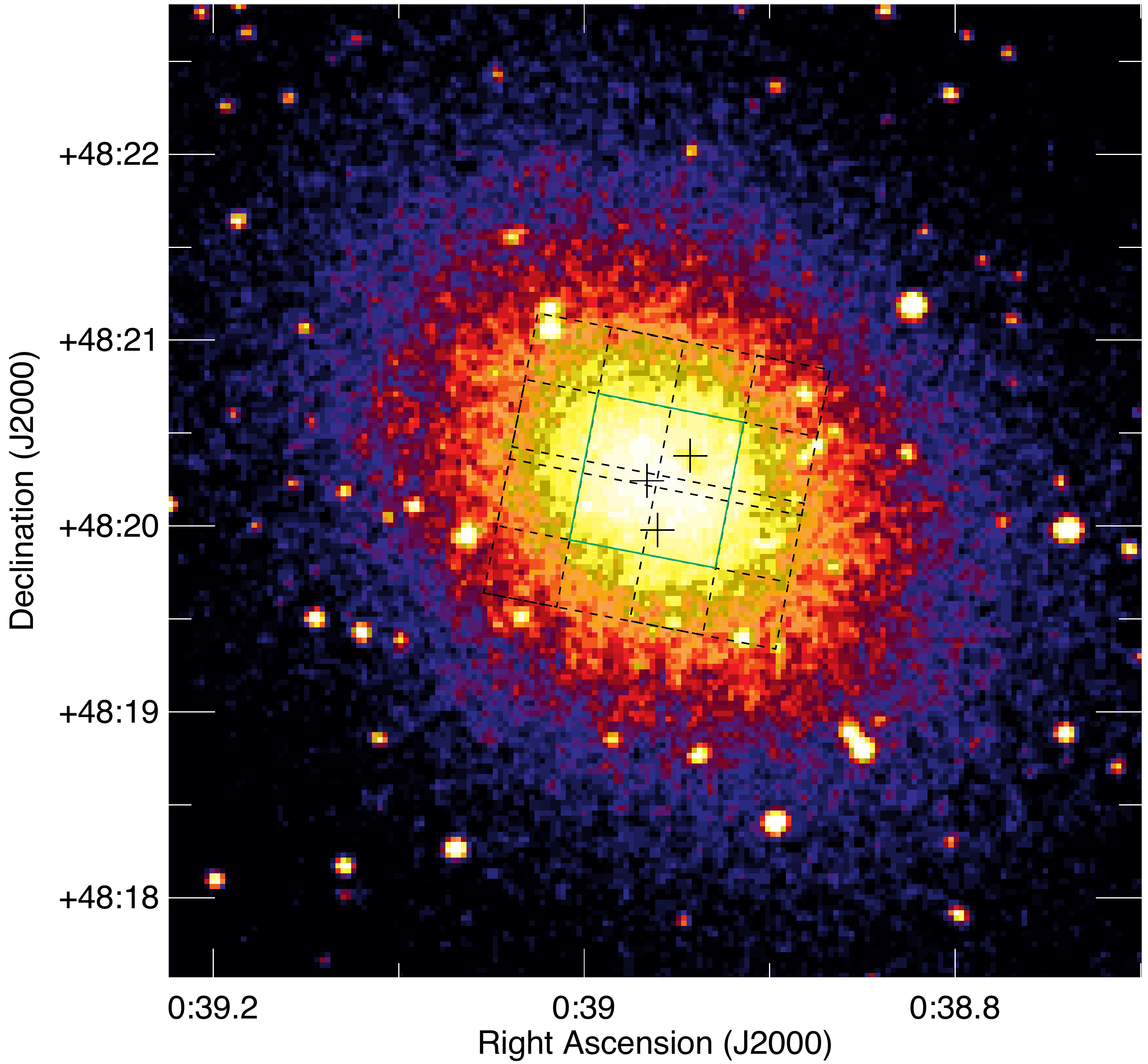}   \\
 \caption{Optical DSS image of NGC\,185 overlaid with the footprint of the \textit{Herschel} PACS [C~{\sc{ii}}] and [O~{\sc{i}}] observations as black dashed and blue solid lines, respectively. The \textit{Spitzer} IRS north, central and south extraction positions from \citet{2010ApJ...713..992M} are indicated as black crosses.} 
              \label{Ima_NGC185_AOR}
\end{figure}

With the PACS spectrometer on board \textit{Herschel} \citep{2010A&A...518L...1P}, we observed $3 \times 3$ raster maps of the fine-structure [C~{\sc{ii}}] 158\,$\mu$m line in a chop-nod observing mode with 2 repetitions, which covers the central $100\arcsec \times 100\arcsec$ area in NGC\,185. The [O~{\sc{i}}] 63\,$\mu$m line was observed only in one raster position covering the central $50\arcsec \times 50\arcsec$ with 8 repetitions (see Fig. \ref{Ima_NGC185_AOR}). The PACS spectra of [C~{\sc{ii}}] (ObsID 1342247543) and [O~{\sc{i}}] (ObsID 1342247544) lines in the central regions of NGC\,185 were taken on July 30th 2012. The data cubes were processed from Level 0 using the telescope normalisation pipeline in \texttt{HIPE} v12.0, with version 65 of the calibration files. After the spectral flat-fielding, the HIPE data cubes were exported to \texttt{PACSman} v.3.5.2 \citep{2012A&A...548A..91L} to perform the line fitting and map projection. The line fitting was done on the full data cloud for each of the 25 spaxels\footnote{Spaxels are spatial pixels that each contain a whole spectrum for a pixel on the sky. For the PACS spectrometer, spaxels have a size of 9.4\,$\arcsec$$\times$9.4\,$\arcsec$.}. The line spectra in each spaxel were modelled with a second-order polynomial and Gaussian function to reproduce the continuum and line emission. The line fit parameters were optimised through a robust non-linear square curve fitting procedure. The spaxels were combined by drizzling to create final intensity maps with pixel size of 3.13$\arcsec$ (i.e., about 1/3 of the size of a single spaxel). The FWHM of the PACS beam at 63\,$\mu$m and 158\,$\mu$m corresponds to $\sim$9.5$\arcsec$ and 11.5$\arcsec$ (or 28\,pc and 32\,pc at the distance of NGC\,185), with a spectral resolution of $\sim$90 km s$^{-1}$, and  $\sim$240 km s$^{-1}$, respectively (see PACS Observer's Manual). The uncertainties inherent to observational noise and line fitting were determined by \texttt{PACSman}. An additional noise factor accounting for a 15$\%$ and 16$\%$ calibration error \citep{2010A&A...518L...2P} at 63\,$\mu$m and 158\,$\mu$m was added in quadrature to the former uncertainties. This calibration uncertainty accounts for the 11-12$\%$ absolute calibration uncertainty (at 63\,$\mu$m and 158\,$\mu$m, respectively) and the 10$\%$ relative uncertainty due to spaxel variations\footnote{http://herschel.esac.esa.int/twiki/pub/Public/PacsCalibrationWeb/\-PacsSpectroscopyPerformanceAndCalibration$\_$v2$\_$4.pdf}.  

\begin{figure*}
\centering
\includegraphics[width=18.5cm]{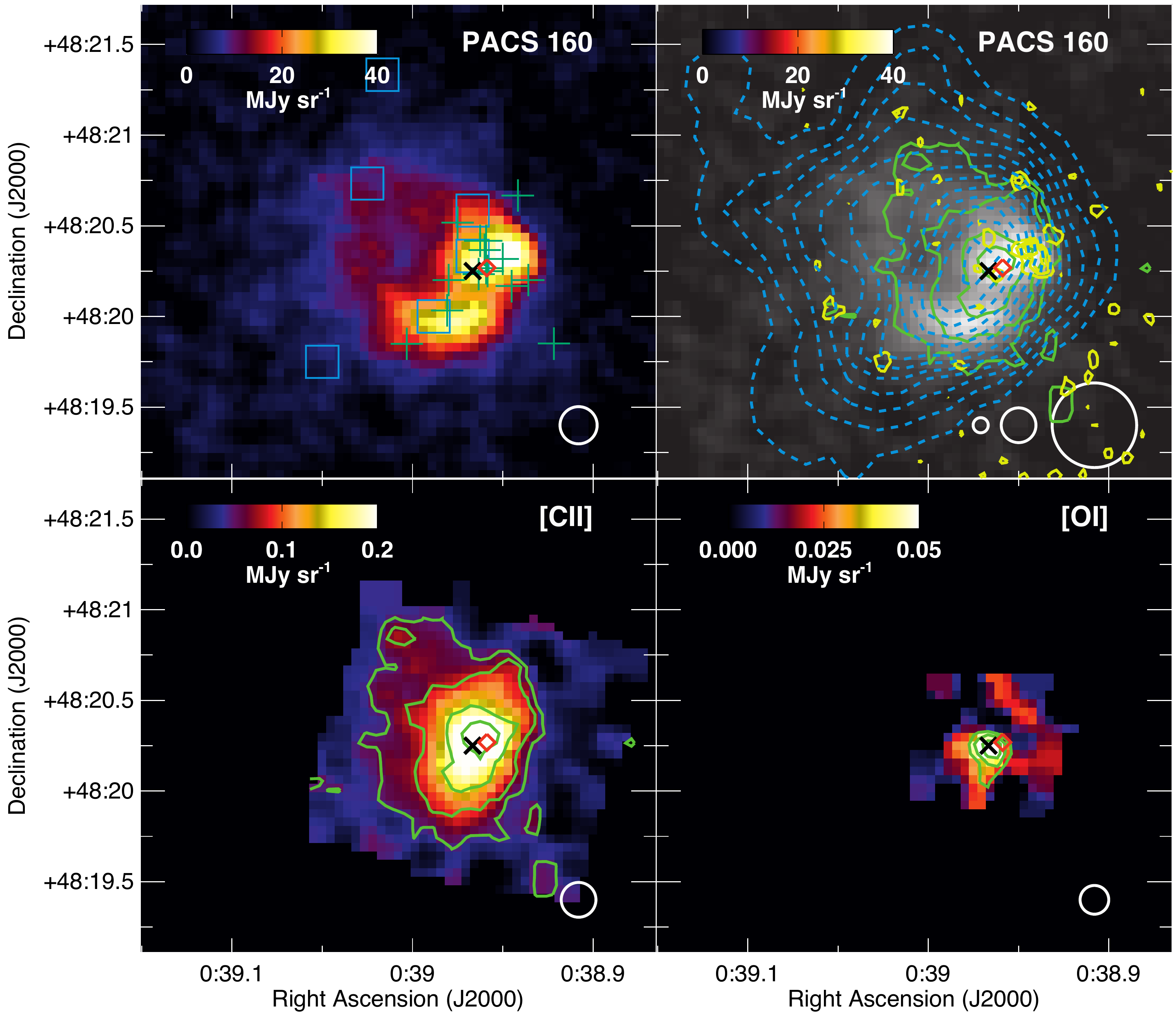}   \\
 \caption{Top panels: \textit{Herschel} PACS\,160\,$\mu$m maps of NGC\,185 with the positions of the blue stars (or star clusters) identified by \citet{1951POMic..10....7B} (green crosses) and clumps in the H {\sc{i}} distribution (cyan squares) overlaid in the left panel. In the right panel, the contours of H {\sc{i}}, CO and [C~{\sc{ii}}] observations are overlaid as blue dashed, yellow and green solid lines, respectively, at their original resolution. The beam sizes of the CO (FWHM$\sim$5$\arcsec$), [C~{\sc{ii}}] (FWHM$\sim$11.5$\arcsec$) and H {\sc{i}} (FWHM$\sim$28$\arcsec$) are indicated in the bottom right corner. The latter figures were taken from \citet{2016MNRAS.459.3900D}. Bottom panels: \textit{Herschel} PACS [C~{\sc{ii}}] (left) and [O~{\sc{i}}] (right) maps of NGC\,185. Beamsizes are indicated in the lower right corner of the respective images. The green contours indicate the 3, 5, 10 and 15$\sigma$ S/N levels for [C~{\sc{ii}}], and the 3, 5 and 8$\sigma$ S/N levels for [O~{\sc{i}}]. The positions of supernova remnant SNR-1 and the centre of NGC\,185 are indicated with a red diamond and black cross, respectively. All images have the same field of view (FOV).}
              \label{Ima_comb}
\end{figure*}

\begin{figure}
\centering
\includegraphics[width=8.5cm]{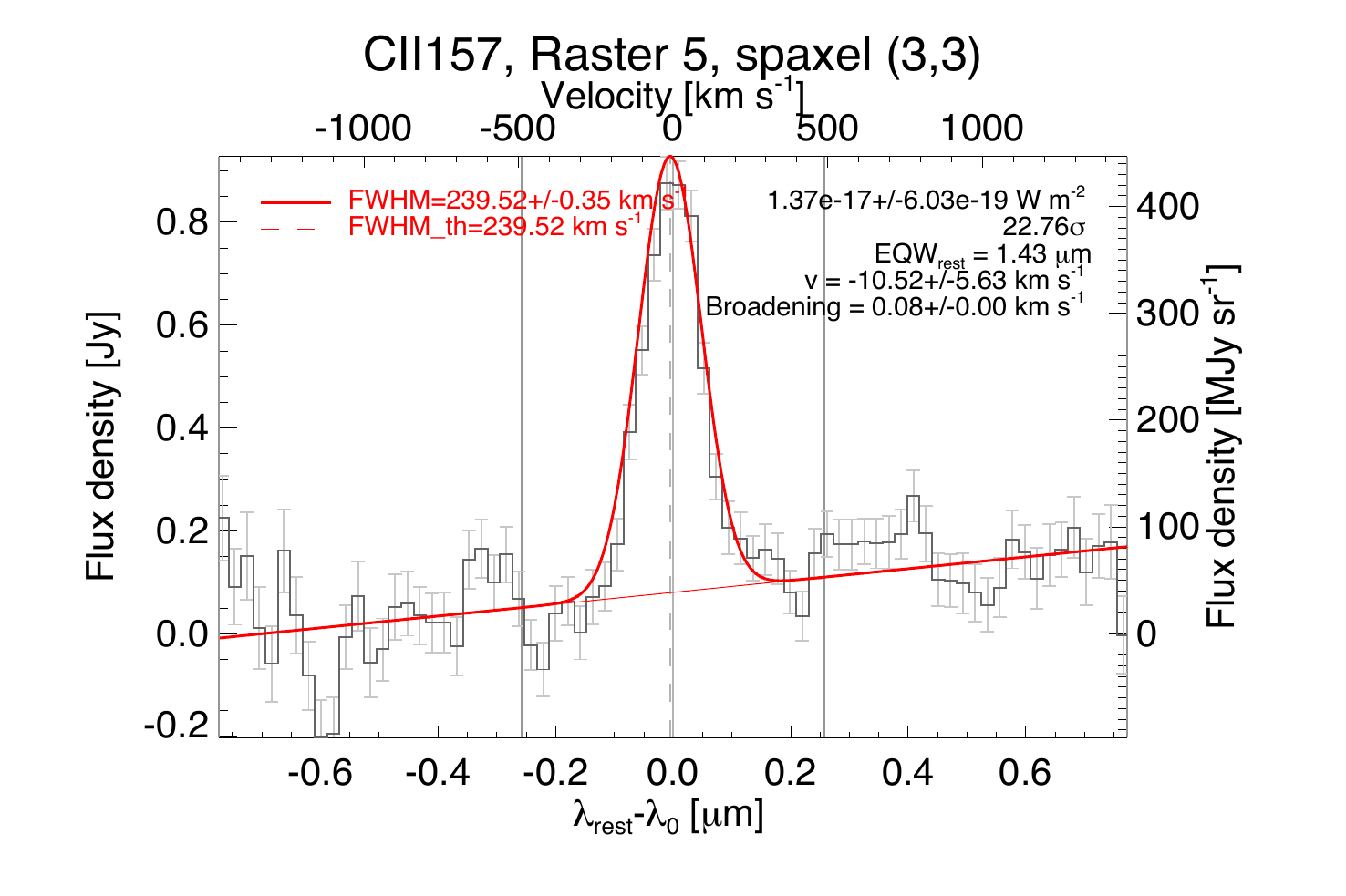}   \\
\includegraphics[width=8.5cm]{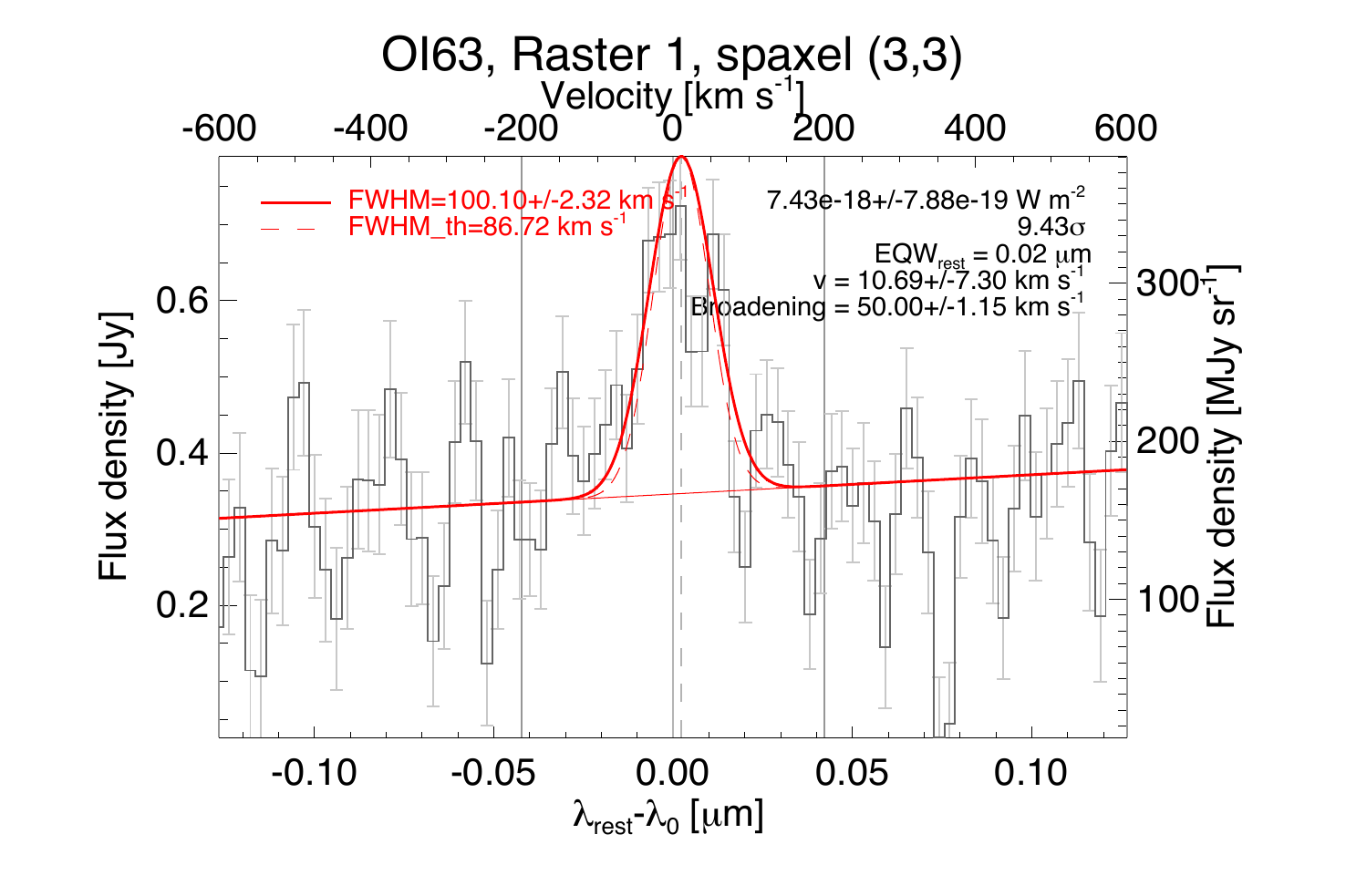}  
 \caption{Representative spectrum of the [C~{\sc{ii}}] 158\,$\mu$m (top) and [O~{\sc{i}}] 63\,$\mu$m (bottom) lines detected in NGC\,185 with the PACS spectrometer in the central spaxel. The thin red lines indicate the first order polynomial that was used to fit the continuum. The thick red lines represent the best fitting Gaussian profile that was fit to the line emission. The dashed red lines indicate the best fit to the data assuming a line width similar to the spectral resolution of the instrument (i.e., 240 km s$^{-1}$ for [C~{\sc{ii}}] and 90 km s$^{-1}$ for [O~{\sc{i}}]).}
              \label{Line_CII_OI}
\end{figure}

[C~{\sc{ii}}] emission is clearly detected from the centre of NGC\,185 (see Fig. \ref{Ima_comb}, bottom left), while a more diffuse component appears to extend towards the east of the galaxy following the distribution of diffuse H~{\sc{i}} and dust clouds (see Fig. \ref{Ima_comb}, top panels). The peak of [C~{\sc{ii}}] is located adjacent to the most massive dust and molecular gas clouds, and coincides with the position of young stars emerging from star forming regions (see Fig. \ref{Ima_comb}, bottom left). This suggests that the majority of [C~{\sc{ii}}] emission arises from photo-dissociation regions positioned in between the star-forming regions and molecular gas reservoirs. 
The [O~{\sc{i}}] line is only detected in the very central region of NGC\,185 (see Fig. \ref{Ima_comb}, bottom right), coinciding with the peak in [C~{\sc{ii}}] emission, and near the location of several young stars. Figure \ref{Line_CII_OI} shows the [C~{\sc{ii}}] 158\,$\mu$m (top) and [O~{\sc{i}}] 63\,$\mu$m (bottom) line profiles detected in the central spaxels of the PACS spectrometer.

\subsection{NRO 45m CO(1-0) observations of NGC\,205}
\label{NRO.sec}
\begin{figure}
\centering
\includegraphics[width=8.5cm]{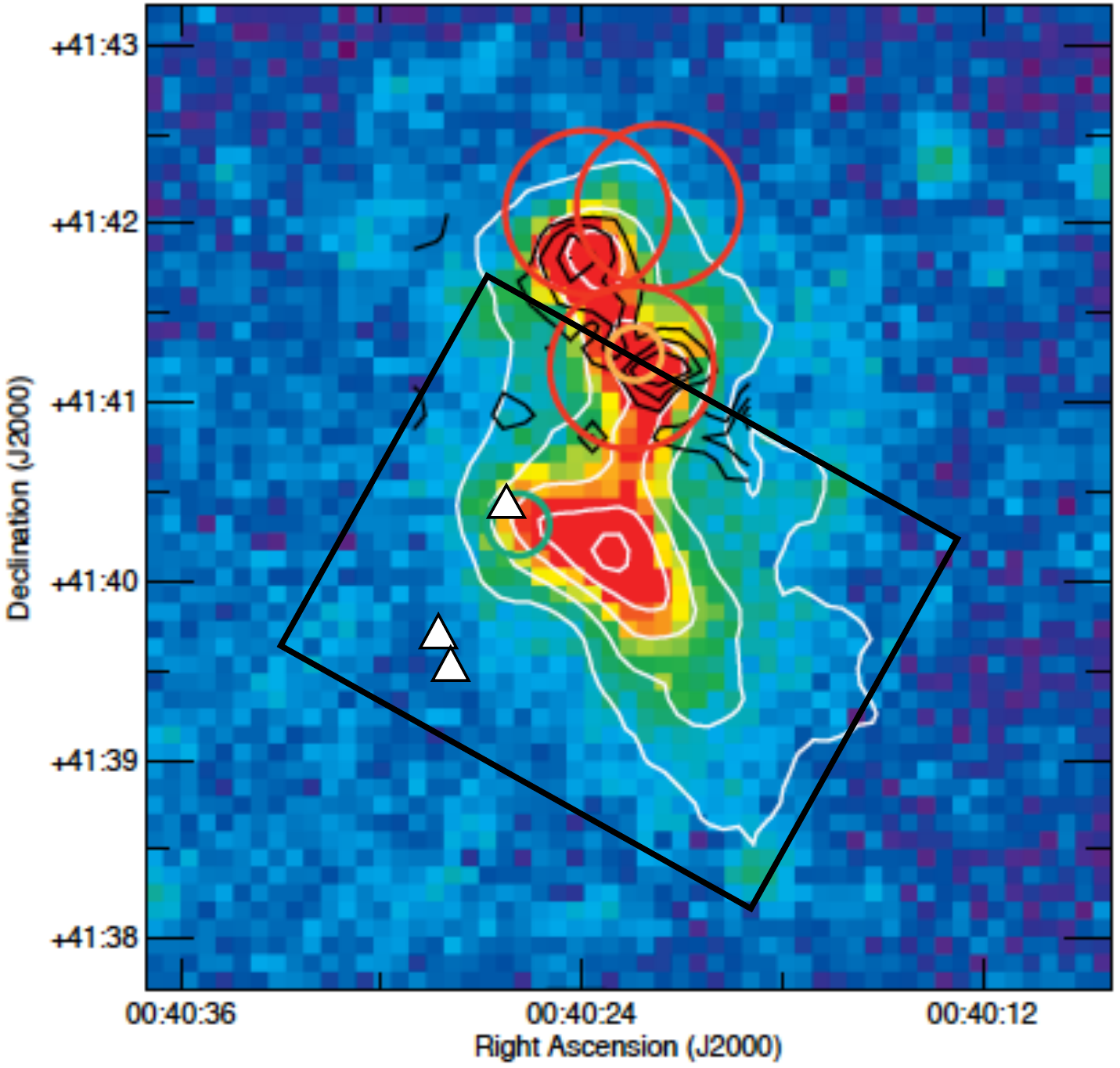}   \\
 \caption{SPIRE 250\,$\mu$m image of NGC\,205 overlaid with H~{\sc{i}} column density contours (\citealt{1997ApJ...476..127Y}, white solid lines) ranging from 2.6 $\times$ 10$^{20}$ cm$^{-2}$ to 1.9 $\times$ 10$^{21}$ cm$^{-2}$, H$_{\text{2}}$ column density contours (\citealt{2012MNRAS.423.2359D}, black solid lines) ranging from 2.6 $\times$ 10$^{20}$ cm$^{-2}$ to 1.9 $\times$ 10$^{21}$ cm$^{-2}$ in steps of 3.3 $\times$ 10$^{20}$ cm$^{-2}$. The red and green circles indicate the CO(1-0) pointings from \citet{1998ApJ...499..209W} and \citet{1996ApJ...464L..59Y}, respectively. The orange circle corresponds to the FWHM (18$\arcsec$) of the observed position with the JCMT at 1.1 mm \citep{1991ApJ...374L..17F}. The black box indicates the region covered by our NRO 45m observations. The white triangles indicate Position 1 (top) and Positions 2A and 2B (bottom) where CO(1-0) line emission is detected from our new NRO\,45\,m data. This Figure was taken from \citet{2012MNRAS.423.2359D}, and updated to include our recent CO observations.} 
              \label{Ima_NGC205_AOR}
\end{figure}

We observed the CO(1-0) line transition with the Nobeyama Radio Observatory (NRO) 45m telescope mapping the central and southern regions of NGC\,205 (see Fig. \ref{Ima_NGC205_AOR}). We mapped a 3.2$\arcmin$ $\times$ 2.7$\arcmin$ region with the On-The-Fly (OTF) mapping mode \citep{2008PASJ...60..445S} with a separation between scans of 5$\arcsec$. The observations were conducted during two separate runs in 2012, extending from January 22nd until January 28th and April 16th until April 25th. During both observing runs, the source IRC+30021 was used for pointing. The average wind speed during both observing runs was less than 5 m s$^{-1}$ on average. The pointing was checked every hour, on average, and found to be accurate within 5$\arcsec$. The full-width at half-maximum (FWHM) of the NRO 45m beam at the CO(1-0) rest frequency of 115 GHz is 16$\arcsec$ (which corresponds to about 64 pc at the distance of NGC\,205).

We observed the CO(1-0) line with the two sideband-separating (2SB) receivers (T100H and T100V) as front-end receivers \citep{2008PASJ...60..435N}.
The analog signal from T100 was converted to 4-8 GHz before being transferred to the digital FX-type spectrometer SAM45 (Spectral Analysis Machine for the 45m telescope). The back-end SAM45 was used with a frequency resolution of 488.24\,kHz which was rebinned to a spectral resolution of 1\,MHz or 2.6 km s$^{-1}$. The typical system noise temperature during the observations ranged between $T_{\text{sys}}$ $\sim$ 200\,K and 300\,K, depending on the weather conditions. The total observing time during the different observing runs was 39 hours, with a total on-source time of 20 hours. 

Data reduction was done with NOSTAR, which is a reduction tool for OTF observations developed by NRO. First of all, the data with pointing errors greater than 5$\arcsec$ were flagged and not used for the construction of the final map. Secondly, the image rejection ratio and the main beam efficiency were used to determine the absolute flux calibration following the method by \citet{2001Kerr}. The uncertainty on the flux calibration is less than 15$\%$, which is based on the combined uncertainty from the measurement of the main beam efficiency and the daily variation of the image rejection ratio ($\sim$5$\%$ during both observing runs). The antenna temperature ($T_{\text{A}}^{*}$) was converted to a main beam temperature ($T_{\text{mb}}$) using a main beam efficiency of $\eta_{\text{mb}}$ = 0.30-0.33\footnote{At the time of the observations, the main beam efficiency was 0.33$\pm$0.03 and 0.30$\pm$0.02 for the T100H and T100V receivers, respectively.} and $T_{\text{mb}}$ = $T_{\text{A}}^{*}$/$\eta_{\text{mb}}$. The final data cube with a grid spacing of 7.5$\arcsec$ was created by convolving with a Gaussian-tapered Bessel function:
\begin{equation}
\frac{J_{1} (r/a)}{r/a} \times \exp \left[-\left( \frac{r}{b} \right)^{2} \right],    \text{with}~a~=~1.55/\pi~and~b~=~2.52,
\end{equation}
with, $J_{1}$, the first order Bessel function and, $r$, the distance between the data and grid point in a pixel. After convolution, the maps have an effective angular resolution of 19.3$\arcsec$ (or 77\,pc at the distance of NGC\,205). Across our map, the average rms sensitivity ranged between $T_{\text{rms}}$ $\sim$ 15 mK and 20 mK at a velocity resolution of 2.6 km s$^{-1}$. 

\begin{figure}
\centering
\includegraphics[width=8.5cm]{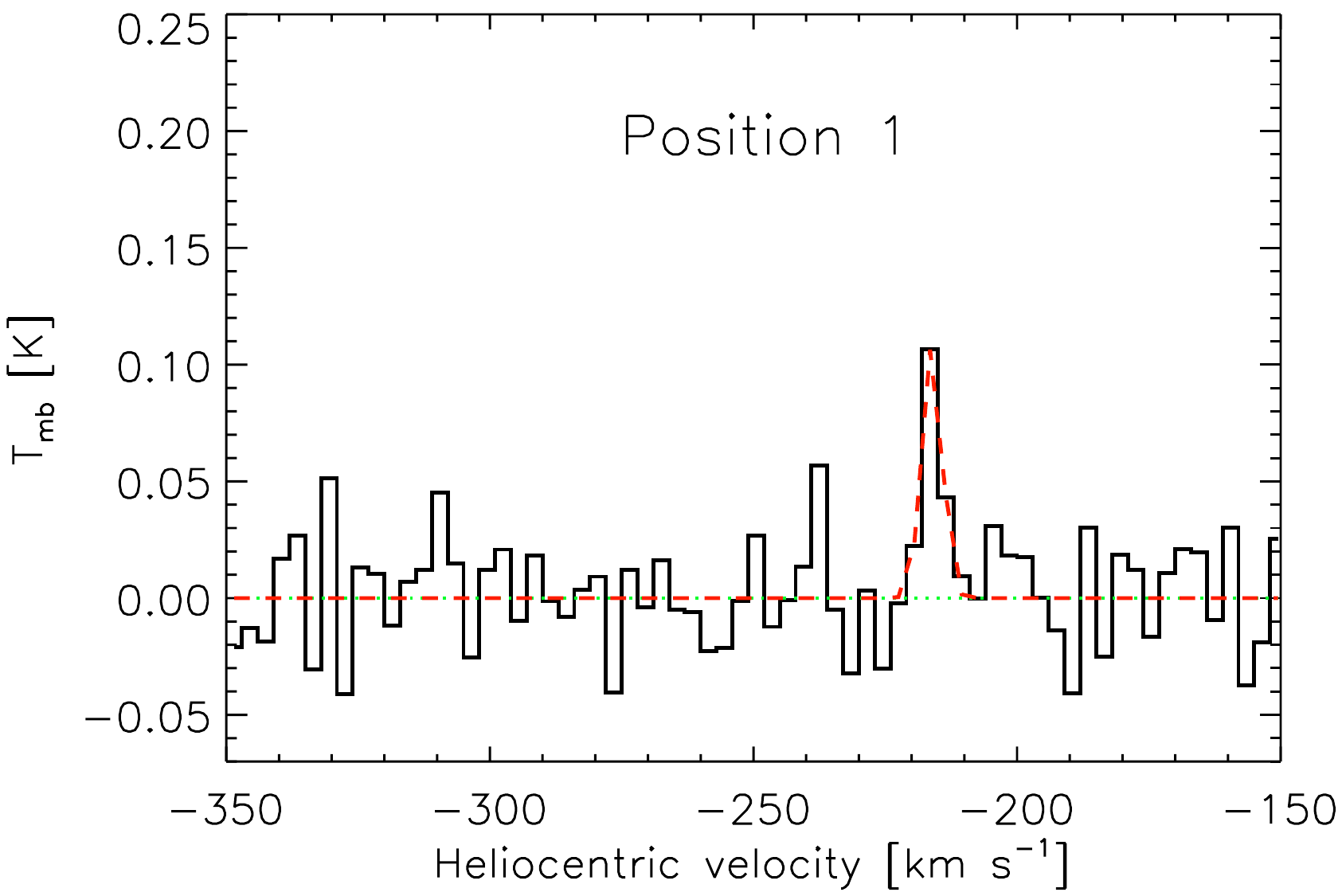}   \\
\includegraphics[width=8.5cm]{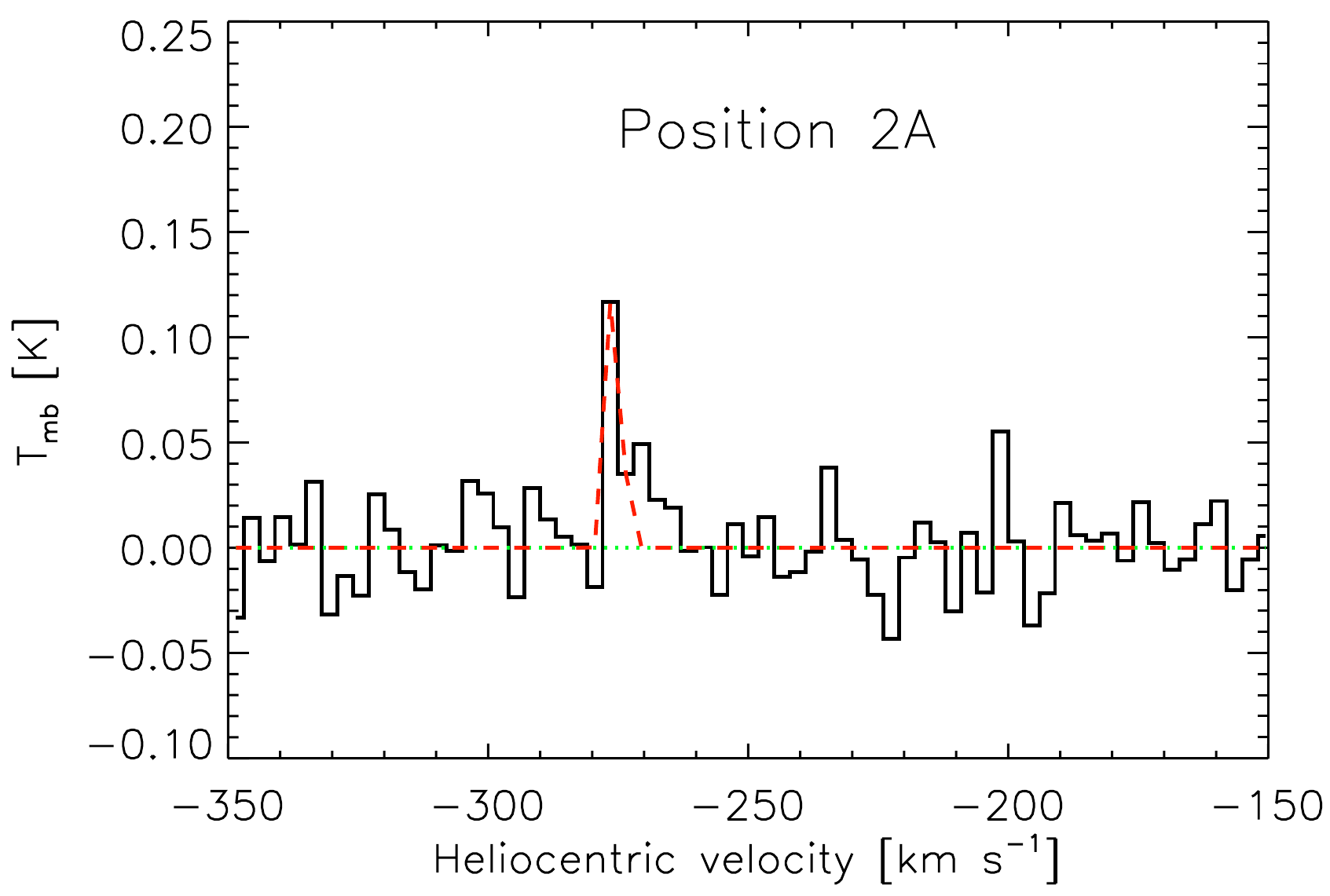}   \\
\includegraphics[width=8.5cm]{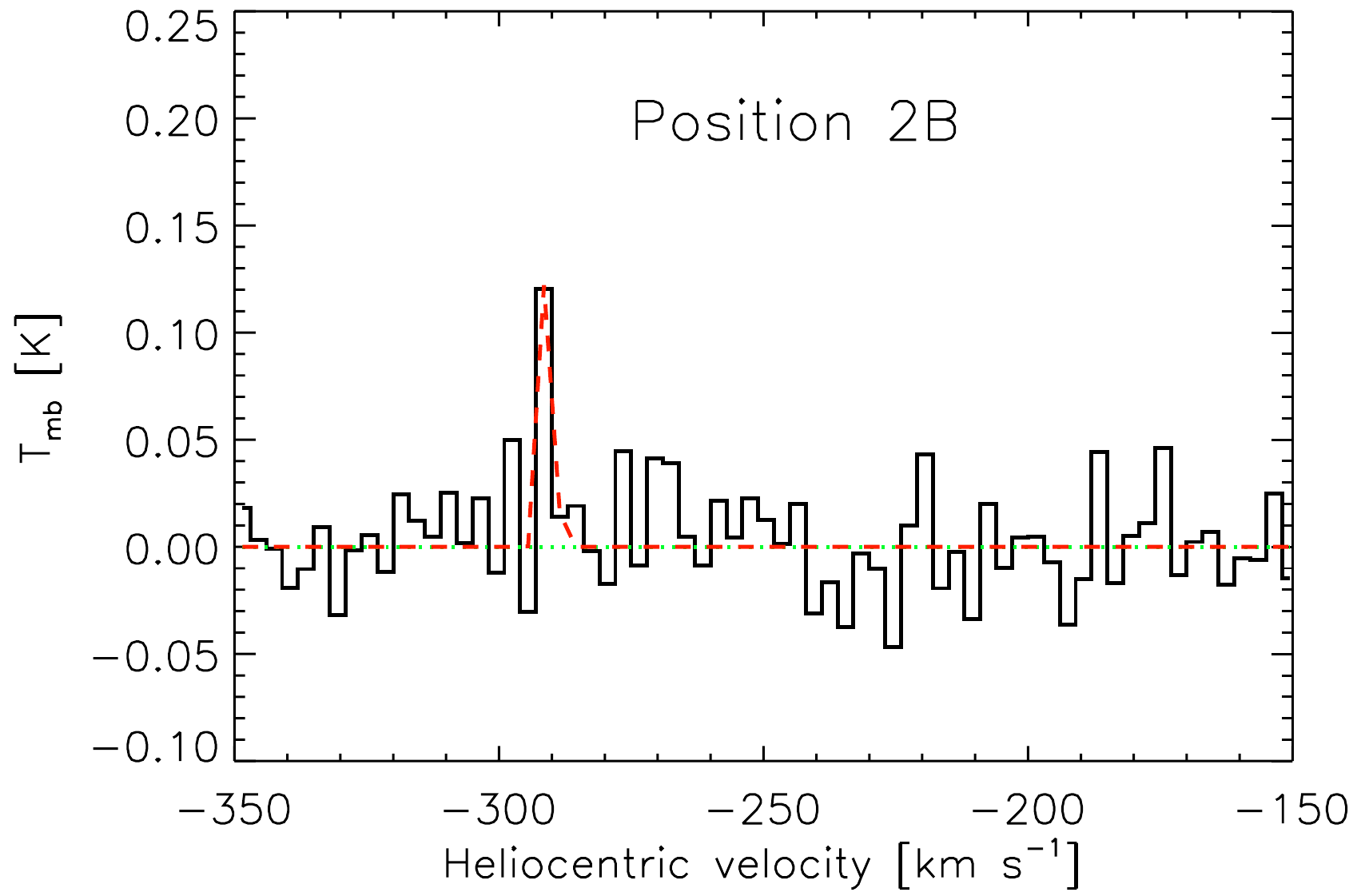}   
 \caption{The CO(1-0) line profiles at the three positions (showing the average emission in a single pixel) where CO(1-0) line emission was detected in our NRO\,45m map covering the southern part of the galaxy NGC\,205.} 
              \label{COlinefit}
\end{figure}

We detect CO(1-0) line emission in three different positions. The line detections in Positions 1 and 2A and 2B are shown in Figure \ref{COlinefit}. The line emission detected in Positions 2A and 2B is separated by only $\sim$15$\arcsec$ (similar to the size of the NRO\,45m beam at 115\,GHz) and with peak velocities that are only 15 km s$^{-1}$ apart, we can not rule out that the two detections originate from the same cloud complex. We average the spectra of adjacent pixels with CO(1-0) detections to derive an average main beam temperature. We fit the baseline of the averaged spectra with a first order polynomial, while the line emission is fit with a Gaussian profile. Table \ref{COlinefit} gives an overview of the equatorial coordinates, central velocity $V$, line width $FWHM$, average main beam temperature $T_{\text{mb}}$ and integrated line intensity $I_{\text{CO}}$ of the three positions. 

The detected CO(1-0) emission at Position 1 is located near the IRAM CO(1-0) detection (with a $\sim$3$\sigma$ significance) reported by \citet{1996ApJ...464L..59Y} at a position (RA, DEC) = (0$^{h}$40$^{m}$25.9$^{s}$, +41$^{\circ}$40$\arcmin$19$\arcsec$). While the central velocity (-218.8 $\pm$ 1.8 km s$^{-1}$) reported by \citet{1996ApJ...464L..59Y} is in fair agreement with our NRO 45m CO(1-0) observations (-216.1$\pm$0.6), we find a smaller line width (4.5 $\pm$ 1.1 km s$^{-1}$) compared to the values reported by \citet{1996ApJ...464L..59Y} (FWHM=19.8$\pm$3.7 km s$^{-1}$). This difference between both CO(1-0) observations can likely be attributed to the IRAM 30m beam (FWHM $\sim$ 21$\arcsec$) being offset by about 7$\arcsec$ from the peak CO(1-0) line emission, which might not have picked up the peak of line emission. The central velocity (-215 km s$^{-1}$) and line width ($\sim$10 km s$^{-1}$) observed for H~{\sc{i}} clouds at the same position are in good agreement with the NRO 45m CO(1-0) observations. Also the line width (6.6$\pm$0.7 km s$^{-1}$) of the CO emission detected with IRAM in the north of NGC\,205 is more consistent with the line width measurement from our NRO 45m observations.

The CO(1-0) emission detected at Position 2A has an average line intensity similar to the line emission detected at Position 1 and is centred around a heliocentric velocity of -275.5 km s$^{-1}$, with a line width (2.5$\pm$0.6 km s$^{-1}$) that is similar to our spectral resolution. The adjacent CO(1-0) detection at Position 2B is centred around a heliocentric velocity of -290.8 km s$^{-1}$ and has a similar narrow line width of 2.4$\pm$0.9 km s$^{-1}$. The central velocities of these CO clouds are at the limit of the stellar velocities (ranging between -~280 and -~140 km s$^{-1}$, \citealt{2010ApJ...711..361G}), and outside of the H~{\sc{i}} velocity range (-260 to -140 km s$^{-1}$) for NGC\,205 \citep{1997ApJ...476..127Y}. We note that the central velocity of the cloud is far from the range of H~{\sc{i}} velocities detected in our Galaxy (-130 to 45 km s$^{-1}$, \citealt{2009ApJ...695..937B}) and this emission does not belong to our Galaxy. The disturbed nature of the distribution of atomic and molecular clouds in NGC\,205 \citep{1997ApJ...476..127Y} and the offset from the main stellar body in NGC\,205, suggests that the gas clouds have not yet settled into a stable configuration. The irregular disposition of this molecular gas cloud might be the result of a recent tidal interaction which has disturbed the gas distribution in NGC\,205. Alternatively, the CO(1-0) line emission at Position 2B might be a false detection and rather correspond to a noise peak, given the small line width which resembles the spectral resolution of the observations. The line emission detected in Position 2A is unlikely to correspond to instrumental noise given its detection in several adjacent pixels (corresponding to 1.5$\times$FWHM), although we can not entirely rule out that it corresponds to a local noise peak.

The line emission in our NRO 45m map of NGC\,205 is detected in only 2 to 7 adjacent 7.5$\arcsec$$\times$7.5$\arcsec$ sized pixels, indicating that the size of molecular clouds in NGC\,205 is extremely small with typical values of $\lesssim$20-25$\arcsec$ (or 80-100\,pc). Interferometric CO(1-0) observations with the BIMA array (with a 40 pc $\times$ 20 pc beam) can barely resolve molecular clouds and measure a cloud size around 40-60 pc for a giant molecular cloud (GMC) in the north of NGC\,205 \citep{1996ApJ...464L..59Y}. Based on a cloud size of 40-60\,pc, the first Larson scaling relation between a cloud's size and velocity dispersion \citep{1981MNRAS.194..809L} predicts a line width of 4.5-5.2 km s$^{-1}$. The similarity with the observed line widths (2.5-4.4 km s$^{-1}$) suggests that the clouds are virialised and experience very little internal gas turbulence. 

We derive molecular gas masses $M_{\text{H}_{2}}$ = $A$ $N_{\text{H}_{2}}$ $m_{\text{H}_{2}}$ where, $A$, is the surface of the CO-emitting region (cf. the number of detected pixels in every position indicated in Table \ref{COdata}), $m_{\text{H}_{2}}$, is the molecular hydrogen mass and the column density of H$_{2}$ is calculated as:
\begin{equation}
\label{Eq_2}
N_{\text{H}_{2}} = X_{\text{CO}} I_{\text{CO(1-0)}}
\end{equation}
with, $I_{\text{CO(1-0)}}$, the integrated main beam line intensity in units of K km s$^{-1}$ and, $X_{\text{CO}}$, the conversion factor. Since the $X_{\text{CO}}$ scaling factor might depend on metallicity, we apply the usual Galactic scaling factor ($X_{\text{CO}}$=2.0$\times$10$^{20}$cm$^{-2}$[K km s$^{-1}$]$^{-1}$, \citealt{1996A&A...308L..21S,2001ApJ...547..792D,2011ApJ...726...81A}) as well as a $H$-band luminosity-dependent conversion factor ($X_{\text{CO}}$=12.5$\times$10$^{20}$cm$^{-2}$[K km s$^{-1}$]$^{-1}$) following the prescriptions from \citet{2002A&A...385..454B}. The $H$-band luminosity of a galaxy is shown to scale with the abundance of metals, and can be considered as a metallicity-dependent $X_{\text{CO}}$ factor. We derive molecular gas masses of $M_{\text{H}_{2}}$=1.0-6.2$\times$10$^{4}$\,M$_{\odot}$, 0.8-5.0$\times$10$^{4}$\,M$_{\odot}$ and 0.2-1.3$\times$10$^{4}$\,M$_{\odot}$ for Positions 1, 2A and 2B, respectively, within the limits of the two different conversion factors. A comparison with previous CO(1-0) observations of the north and central regions of NGC\,205 (see Section \ref{HIandCO.sec}) show that the molecular gas clouds in the south account for only one tenth of the total molecular gas reservoir in NGC\,205.

Total molecular gas masses (see Table \ref{ISMmasses}) using a metallicity-dependent $X_{\text{CO}}$ factor are more than three times higher compared to the atomic gas mass in NGC\,205 ($M_{\text{HI}}$=4$\times$10$^{5}$\,M$_{\odot}$), which seems unrealistic given the low star formation activity in NGC\,205. The choice of a Galactic $X_{\text{CO}}$ factor is also consistent with the Galactic conversion factor derived by \citet{2008ApJ...686..948B} for a molecular cloud in the centre of NGC\,205 based on virial mass assumptions. Molecular gas mass depletion factors ($\log$ $\tau_{\text{H}_{2}}$ $\sim$ 0.6-0.8\,Gyr) based on $H$-band luminosity-dependent $X_{\text{CO}}$ factors would however better agree with the trend between molecular gas depletion time scale and specific star formation rate (sSFR=SFR/M$_{\star}$) observed for the COLD GASS sample \citep{2011MNRAS.415...61S}. To account for uncertainties on the $X_{\text{CO}}$ factor, we will mention molecular gas masses derived from both Galactic and $H$-band luminosity-dependent $X_{\text{CO}}$ factors in the remainder of this work. 

\begin{table*}
\caption{Overview of the CO line fitting parameters with the position coordinates (of the brightest pixel), average main beam temperature, velocity, line width and integrated CO line intensity for every position. The last two columns present the molecular gas masses derived from the average integrated CO line intensity summed over all detected pixels based on Galactic and metallicity-dependent $X_{\text{CO}}$ factors.}
\label{COdata}
\centering
\begin{tabular}{|lll|ccccccc|}
\hline 
Position & RA (J2000) & DEC (J2000) & $T_{\text{mb}}$ & $v_{\text{cen}}$ & $FWHM$ & $I_{\text{CO}}$ & $N_{\text{pix}}$ & $M_{\text{H}_{2}}$($X_{\text{CO,Gal}}$) & $M_{\text{H}_{2}}$($X_{\text{CO,Met}}$)  \\
 & [$^{h}$ $^{m}$ $^{s}$] & [$^{\circ}$ $\arcmin$ $\arcsec$] &  [K] & [km s$^{-1}$] & [km s$^{-1}$] &  [K km s$^{-1}$] & & [M$_{\odot}$] & [M$_{\odot}$] \\
 \hline
Position 1   & 00:40:26.0 & +41:40:36.4 & 0.11$\pm$0.02 & -216.1$\pm$0.6 & 4.5$\pm$1.1 & 0.49 & 7 & 1.0$\times$10$^{4}$ & 6.2$\times$10$^{4}$ \\
Position 2A & 00:40:27.4 & +41:39:28.9 & 0.18$\pm$0.07 & -275.5$\pm$0.3 & 2.5$\pm$0.6 & 0.46 & 6 & 0.8$\times$10$^{4}$ & 5.0$\times$10$^{4}$ \\ 
Position 2B & 00:40:28.0 & +41:39:43.9 & 0.16$\pm$0.07 & -290.8$\pm$0.5 & 2.4$\pm$0.9 & 0.37 & 2 & 0.2$\times$10$^{4}$ & 1.3$\times$10$^{4}$ \\ 
 \hline
\end{tabular}
\end{table*}

\subsection{SPIRE FTS spectroscopy data of NGC\,205}
\label{SPIREFTS.sec}
\begin{figure}
\centering
\includegraphics[width=8.5cm]{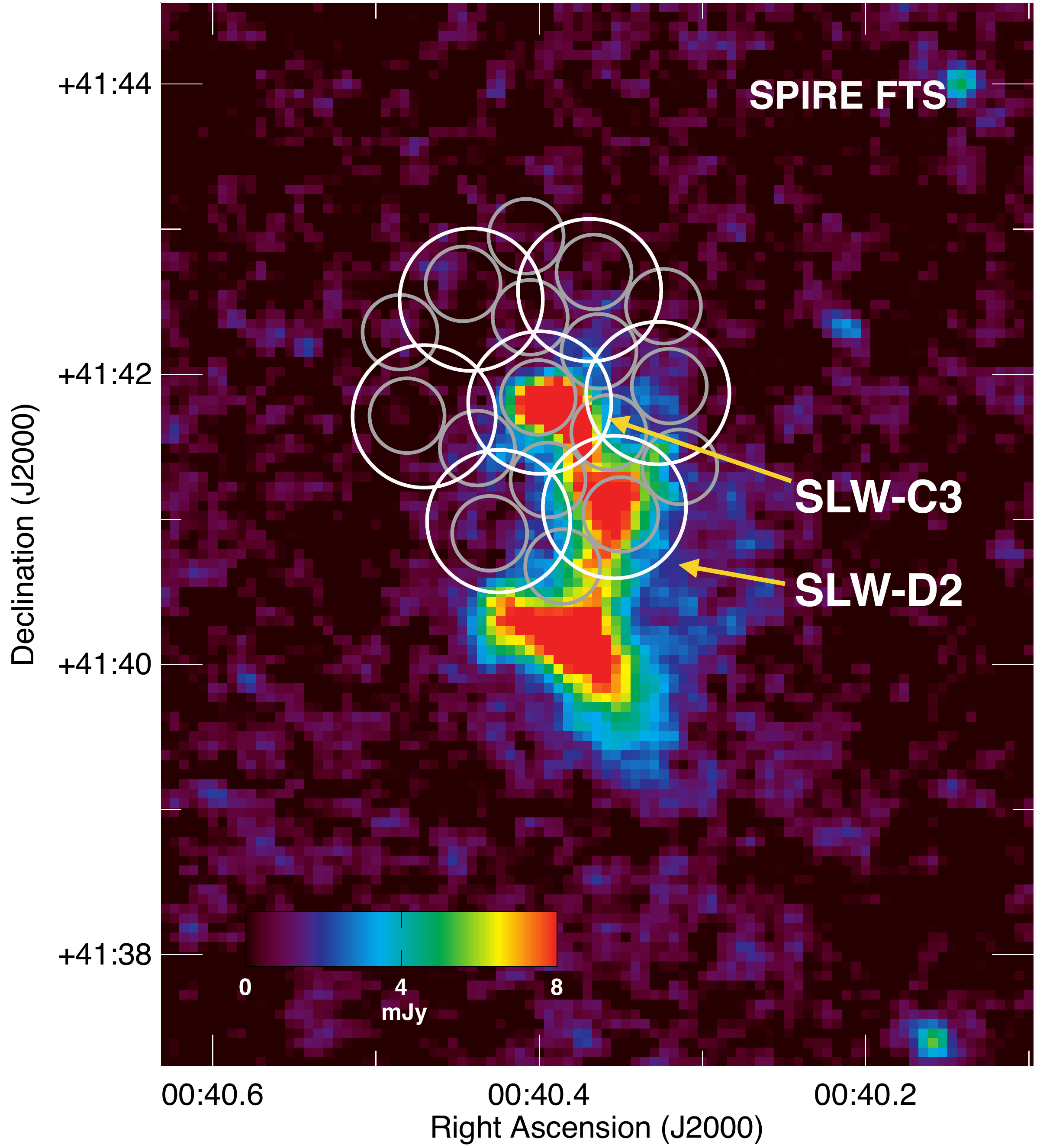}   
 \caption{PACS\,160\,$\mu$m image of NGC\,205 overlaid with the SPIRE FTS pointing targeting the CO peak in the North with individual jiggle positions for the SSW and SLW detectors indicated as grey and white circles, respectively. The SPIRE FTS SLW-C3 and SLW-D2 bolometers are labelled in the Figure.} 
              \label{NGC205_SPIREFTS_AOR}
\end{figure}

\textit{Herschel} observations of NGC\,205 were acquired as part of the Guaranteed Time (GT) program $``$\textit{Very Nearby Galaxies Survey (VNGS)}$"$ (PI: C. Wilson). The SPIRE FTS spectra were obtained in sparse spatial sampling and high spectral resolution mode, covering the 194-671\,$\mu$m wavelength range. One single pointing targeting the CO peak in the North of NGC\,205 (see Fig. \ref{NGC205_SPIREFTS_AOR}) was observed with 74 repetitions. The 35 detectors of the SSW (SPIRE Short Wavelength) array covered the 194-313\,$\mu$m range, while the SLW (SPIRE Long Wavelength) array of 19 detectors covered the 303-671\,$\mu$m wavelength range. The SSW and SLW arrays have an average FWHM of 19$\arcsec$ and 34$\arcsec$, respectively \citep{2013ApOpt..52.3864M}. 

The SPIRE FTS data were reduced in \texttt{HIPE} v14.0.0, with version SPIRE$\_$CAL$\_$14$\_$2 of the calibration files. The standard pipeline in \texttt{HIPE} for single pointing SPIRE spectrometer observations was used for data reduction, assuming a point source calibration without apodisation. The standard pipeline included a first and second order deglitching procedure, non-linearity and phase corrections, baseline subtraction, and corrections for the telescope and instrument emission. We fit spectral lines in the SPIRE FTS data using the SPIRE Spectrometer Line Fitting algorithm in \texttt{HIPE}. All emission lines of interest (i.e., CO line transitions from $^{12}$CO(4-3) up to $^{12}$CO(13-12) and the two [C~{\sc{i}}] line and [N~{\sc{ii}}]\,205\,$\mu$m line transitions) were fit simultaneously using a third order polynomial for the continuum while the line profile is fit with a sinc function. We assume the line profiles have a width that corresponds to the spectral resolution of the instrument (ranging from 280 to 970 km s$^{-1}$ towards longer wavelengths). Based on the typical line widths ($\lesssim$20\,km s$^{-1}$, \citealt{1996ApJ...464L..59Y,1997ApJ...476..127Y}) for the observed H {\sc{i}} and CO line transitions in NGC\,205, we are confident that the line profiles will be set by the instrument's spectral imprint. 

We do not detect CO, [C~{\sc{i}}] or [N~{\sc{ii}}] emission in any of the SPIRE FTS bolometers. The 1$\sigma$ upper limits derived from the line fitting algorithm for the spectra of the SPIRE SLW-C3 bolometer constrains the [C~{\sc{i}}] integrated line fluxes in the CO-peak in the North ($I_{\text{[CI] 1-0}}$ $\leq$ 1.06$\times$10$^{-18}$ W m$^{-2}$ or 65 Jy km s$^{-1}$; $I_{\text{[CI] 2-1}}$ $\leq$ 1.09$\times$10$^{-18}$ W m$^{-2}$ or 40\,Jy km s$^{-1}$) and the spectra of the SPIRE SLW-D2 bolometer similarly constrain [C~{\sc{i}}] integrated line fluxes in the centre of NGC\,205 ($I_{\text{[CI] 1-0}}$ $\leq$ 1.04$\times$10$^{-18}$ W m$^{-2}$ or 63\,Jy km s$^{-1}$; $I_{\text{[CI] 2-1}}$ $\leq$ 1.07$\times$10$^{-18}$ W m$^{-2}$ or 40\,Jy km s$^{-1}$). We furthermore use the 1$\sigma$ upper limits on the [N~{\sc{ii}}]\,205\,$\mu$m flux ($I_{\text{[NII] 1-0}}$ $\leq$ 8.08$\times$10$^{-19}$ W m$^{-2}$) in the centre of NGC\,205 to constrain the [C~{\sc{ii}}] contribution from ionised gas in Section \ref{Origin.sec}.

\subsection{Ancillary data}
For NGC147, we only have H~{\sc{i}} and CO observations to constrain the gas mass. We rely on the upper gas mass limits from H~{\sc{i}} and H$_{2}$ observations reported by \citet{1997ApJ...476..127Y} and \citet{1998ApJ...507..726S}, respectively, to constrain the gas content in NGC\,147. 

For NGC185, we have H~{\sc{i}}, CO and H$\alpha$ observations and \textit{Spitzer} Infrared Spectrograph (IRS) spectra with rotational $H_{2}$ lines to constrain the atomic, cold molecular, ionised and warm molecular gas masses.
We reduced H~{\sc{i}} data for NGC\,185 observed with the Very Large Array (VLA) in configurations C (8 hr) and D (4.5 hr), in a similar way as presented in \citet{1997ApJ...476..127Y}. We derive a moment-0 map of the H~{\sc{i}} observations following the same strategy as \citet{1997ApJ...476..127Y}. The H~{\sc{i}} map was derived at medium resolution of 28$\arcsec$ $\times$ 26$\arcsec$ (or 84 pc $\times$ 78 pc) with rms noise level of 0.54 mJy beam$^{-1}$ (or 0.46 K), which corresponds to a H~{\sc{i}} column density of $\sim$3 $\times$ 10$^{19}$ cm$^{-2}$. At the medium resolution of 28$\arcsec$ $\times$ 26$\arcsec$, we observe a peak in H~{\sc{i}} column density N$_{\text{HI}}$ $\sim$ 3.1 $\times$ 10$^{20}$ cm$^{-2}$. Assuming optically thin H~{\sc{i}} emission, we derive a total H~{\sc{i}} mass of M$_{\text{HI}}$ = 1.1 $\times$ 10$^{5}$ M$_{\odot}$, scaled to our adopted distance D=0.616 Mpc.
We, furthermore, use the CO map obtained from interferometric observations with the Berkeley-Illinois-Maryland Association (BIMA) array presented by \citet{2001AJ....122.1747Y}. For the analysis in this paper, we use the CO intensity map with 5.5$\arcsec$ $\times$ 4.6$\arcsec$ (17\,pc $\times$ 14\,pc) resolution and a rms noise level of 0.070 Jy beam$^{-1}$ (or 0.25 K).

For NGC\,205, we have H~{\sc{i}}, CO and H$\alpha$ observations to constrain the atomic, cold molecular and ionised gas masses. The CO observations in the north and centre of NGC\,205 \citep{1996ApJ...464L..59Y,1998ApJ...499..209W} are complemented with our new NRO\,45m observations covering the southern regions of NGC\,205 (see Section \ref{NRO.sec}). We furthermore have \textit{Herschel} SPIRE FTS spectra with [C~{\sc{i}}] line transitions that allow us to constrain the CO-dark molecular gas content in NGC\,205. The ancillary H~{\sc{i}} and JCMT CO(3-2) data sets used for NGC\,205 were described in \citet{2012MNRAS.423.2359D}.



\section{Physical gas characteristics}
\subsection{Origin of the [C~{\sc{ii}}] emission}
\label{Origin.sec}
With an ionisation potential of 11.3 eV, the [C~{\sc{ii}}] line emission in NGC\,185 can originate from photo-dissociation regions (PDRs), the cold neutral medium (CNM), and ionised gas phases. The [C~{\sc{ii}}] contribution of the ionised gas phase is considered to be negligible in NGC\,185 given the weak emission of ionised gas tracers (e.g., H$\alpha$) and the absence of a strong radiation field \citep{2016MNRAS.459.3900D}. The latter argument is also supported by the negligible [C~{\sc{ii}}] contribution from ionised gas in NGC\,205 assuming that its ISM conditions are similar to NGC\,185. The [C~{\sc{ii}}] contribution from ionised gas is estimated in NGC\,205 based on the 1$\sigma$ upper limit on its [N~{\sc{ii}}]\,205\,$\mu$m emission as observed with the SPIRE FTS instrument onboard \textit{Herschel} (see Section \ref{SPIREFTS.sec}). Theoretical models predict line ratios of [C~{\sc{ii}}]/[N~{\sc{ii}}]$_{\text{205}}$ $\sim$ 3-4 for a range of different electron densities \citep{2006ApJ...652L.125O}. Based on the lower limit on the observed [C~{\sc{ii}}]/[N~{\sc{ii}}]$_{\text{205}}$ line ratio ([C~{\sc{ii}}]/[N~{\sc{ii}}]$_{\text{205}}$ $\geq$123) in NGC\,205, we estimate a maximum [C~{\sc{ii}}] contribution of the ionised gas phase of $\leq$4\,$\%$. To predict the [C~{\sc{ii}}] emission produced in the cold neutral medium excited by collisions with hydrogen atoms or molecules, we apply Eq. 1 from \citet{1997ApJ...483..200M}:
\begin{equation}
\label{EqMadden}
I_{\text{[CII]}} = 2.35 \times 10^{-21} N_{\text{C+}} \left( \frac{2 \times \exp(-91.3/T)}{1 + 2 \times \exp(-91.3/T) + (n_{\text{crit}}/n_{\text{H}})} \right),
\end{equation}
with, $I_{\text{[CII]}}$, the [C~{\sc{ii}}] intensity in units of erg s$^{-1}$ cm$^{-2}$ sr$^{-2}$ and, $N_{\text{C+}}$, the C$^{+}$ column density calculated as $N_{\text{C+}}$ =  $X_{\text{C+}}$ $N_{\text{HI}}$ with, $N_{\text{HI}}$, the H~{\sc{i}} column density. We assume a gas temperature $T$ = 50K and gas density $n_{\text{H}}$ = 100 cm$^{-3}$ typical for the cold neutral medium \citep{1997ApJ...483..200M}. The $C^{+}$/$H$ abundance ($X_{C^{+}}$ $\sim$ 5.0 $\times$ $10^{-5}$) is scaled relative to the solar carbon abundance ($X_{C}$ $\sim$ 1.4 $\times$ $10^{-4}$, \citealt{1997ApJ...482L.105S}) based on the metallicity of NGC\,185 (0.36 Z$_{\odot}$, see Section \ref{Metal.sec}) under the assumption\footnote{This assumption might not be appropriate given the soft radiation field and low gas temperature $T \sim$40-70\,K in NGC\,185, which might prevent the ionisation of carbon and/or excitation of C$^{+}$. For lower C$^{+}$ abundances, the CNM contribution to the [C~{\sc{ii}}] emission will be smaller than the values quoted here.} that all of the carbon is in the form of C$^{+}$. We assume a critical density $n_{\text{crit}}$(H) $\sim$ 1.6 $\times$ $10^{3}$ cm$^{-3}$ \citep{2012ApJS..203...13G} for collisions with H atoms.

We convolve the observed [C~{\sc{ii}}] map to the resolution of the H~{\sc{i}} observations ($\sim$ 28$\arcsec$) to compare the observed [C~{\sc{ii}}] intensity with the predicted contribution from the cold neutral medium. We find [C~{\sc{ii}}] contributions ranging from 12$\%$ to 25$\%$, implying that only up to a quarter of the [C~{\sc{ii}}] emission originates from the cold neutral medium. The highest contributions from H~{\sc{i}} clouds to the [C~{\sc{ii}}] emission occur in the more diffuse emission regions, while the [C~{\sc{ii}}] contribution from the cold neutral medium reaches a minimum in the dust mass peak south-west of the centre. The small contribution from H~{\sc{i}} clouds suggests that the majority of [C~{\sc{ii}}] emission in NGC\,185 originates from PDRs.

Based on the total integrated CO flux density reported by \citet{2001AJ....122.1747Y} ($S_{\text{CO}}$ $\sim$ 8.8 Jy km s$^{-1}$) and the sum of the $\geq$ 3$\sigma$ regions in the [C~{\sc{ii}}] map ($F_{\text{[CII]}}$(total)~=~1.33$\times$10$^{-16}$~W~m$^{-2}$), we derive a [C~{\sc{ii}}]/CO ratio of $\sim$ 3.9$\times$10$^{3}$. The line ratio in NGC\,185 is higher compared to the [C~{\sc{ii}}]/CO ratio observed in NGC\,205 ($\sim$1.9$\times$10$^{3}$, \citealt{2012MNRAS.423.2359D}), similar to the range of values observed in starburst galaxies \citep{1991ApJ...373..423S,2001A&A...375..566N}, but at the low end of the line ratios observed in low-metallicity star-forming dwarf galaxies, ranging from 4,000 to 80,000 \citep{2000NewAR..44..249M,2010A&A...518L..57C,2016arXiv160304674M}. In these low-metallicity star-forming dwarf galaxies, CO molecules are more easily photo-dissociated due to the hard radiation and a porous ISM structure, leaving behind a layer of self-shielding H$_{2}$ that is not traced by CO observations. Based on the low [C~{\sc{ii}}]/CO line ratios in dSphs, this CO-dark gas component \citep{2010ApJ...716.1191W} is expected to be significantly less important in NGC\,185 and NGC\,205 compared to the CO-dark gas reservoir in metal-poor star-forming dwarfs. The UV radiation fields in these dSphs is several times weaker compared to star-forming dwarf galaxies, enabling CO molecules to survive and trace the bulk of H$_{\text{2}}$ gas mass.

\subsection{Photoelectric efficiency}
\label{PE.sec}
A map of the total-infrared (TIR) emission in NGC\,185 is calculated based on the MIPS\,24\,$\mu$m, PACS\,100\,$\mu$m and PACS\,160\,$\mu$m maps and the prescriptions of \citet{2013MNRAS.431.1956G}. For the computation of the TIR emission, all maps have been convolved to the resolution of the PACS\,160\,$\mu$m waveband with the appropriate kernels from \citet{2011PASP..123.1218A}. To measure line intensities, we similarly convolve all line maps to the resolution of the PACS\,160\,$\mu$m waveband. Since the [O~{\sc{i}}] line is only detected in the centre of NGC\,185, we measure fluxes ($F_{\text{[CII]}}$ = 3.5$\pm$0.3$\times$10$^{-17}$ W m$^{-2}$, $F_{\text{[OI]}}$ = 8.6$\pm$5.0$\times$10$^{-18}$ W m$^{-2}$, $F_{\text{TIR}}$ = 2.4$\pm$0.1$\times$10$^{-15}$ W m$^{-2}$)\footnote{The uncertainty on the TIR emission accounts for the observational uncertainties on the MIPS\,24\,$\mu$m and PACS measurements and the scatter in the calibrations of \citet{2013MNRAS.431.1956G}.} within a circular aperture of radius R=12.1$\arcsec$ (or similar to the FWHM of the PACS\,160\,$\mu$m beam) towards the centre of NGC\,185. Since [O~{\sc{i}}] emission is only detected in the central spaxel, we can also measure the total [O~{\sc{i}}] flux ($F_{\text{[OI]}}$ = 10.6$\pm$0.8$\times$10$^{-18}$ W m$^{-2}$) by applying a point-source correction to the flux detected in the central spaxel. Given that the uncertainty on the corrected flux measurement from the central spaxel is lower, we will use the latter [O~{\sc{i}}] flux measurement in the remainder of this work.

Based on these line measurements for the central region, we derive [C~{\sc{ii}}]/TIR = 0.015$\pm$0.004 and [C~{\sc{ii}}]+[O~{\sc{i}}]/TIR = 0.021$\pm$0.004 line ratios. The line ratios are indicative of the efficiency of the photoelectric effect in case the [C~{\sc{ii}}] and [O~{\sc{i}}] line emission is a good proxy for the gas cooling (and thus gas heating) and TIR is representative of the energy of stars that goes into heating the dust. With ratios higher than 1$\%$, the photoelectric heating of neutral gas in the central regions of NGC\,185 is considered more efficient compared to the average [C~{\sc{ii}}]/TIR ratios (0.1-1\,$\%$) in normal star-forming galaxies (e.g., \citealt{2001ApJ...561..766M,2008ApJS..178..280B,2016Smith}). Given the soft radiation field in NGC\,185 and the bright features of polycyclic aromatic hydrocarbons (PAHs) detected in the \textit{Spitzer} IRS spectra \citep{2010ApJ...713..992M}, the high photo-electric efficiency might be attributed to a high PAH abundance and/or a low fraction of grain charging. Similarly high photoelectric efficiencies were observed in a sample of low surface brightness dwarf galaxies \citep{2016AJ....151...14C}.

\subsection{PDR diagnostics}
\label{PDR.sec}
To examine the state of the gas in NGC\,185, the observed [C~{\sc{ii}}] and [O~{\sc{i}}] line and total-infrared (TIR) emission are compared to PDR models using the PDR Toolbox (PDRT, \citealt{2008ASPC..394..654P}). For a comparison of the observed [O~{\sc{i}}]/[C~{\sc{ii}}] and ([C~{\sc{ii}}]+[O~{\sc{i}}]/TIR) line ratios to PDR models, we assume that all of the [C~{\sc{ii}}] and [O~{\sc{i}}] emission originates from PDRs. The line ratios in the PDR toolbox are calculated for a plane-parallel geometry with elemental abundances and grain properties fixed for a metallicity 1$Z_{\odot}$\footnote{Although the metallicity of NGC\,185 is sub-solar, we assume to first order that the effect of metallicity on the observed line ratios is negligible.}. Due to the assumption in the PDR models of a slab geometry that is illuminated and emitting on one side, we need to make certain corrections to the observed line emission before comparing it to PDR models. We apply a similar strategy to correct our observations as followed by \citet{2013ApJ...776...65P,2014ApJ...787...16P} and \citet{2015A&A...575A..17H}. Since we observe the front and back side emission of clouds in NGC\,185 (under the assumption of optically thin infrared emission), we divide the observed TIR emission by a factor of two to be consistent with the model that only accounts for emission from the front side of the cloud. Because the [O~{\sc{i}}] line becomes optically thick relatively fast, we multiply the observed [O~{\sc{i}}] line emission by a factor of two to account for the clouds that have their optically thick side oriented towards us.

Based on the corrected line ratios ([O~{\sc{i}}]/[C~{\sc{ii}}] = 0.61 $\pm$ 0.15 and ([C~{\sc{ii}}]+[{O~{\sc{i}}}])/TIR = 0.047 $\pm$ 0.009), PDR models predict a ISRF scaling factor, $G_{0}$ = 31.6, and a hydrogen gas density, $n_{\text{H}}$ = 10$^{3.75}$ cm$^{-3}$. The PDR line diagnostics suggest a stronger radiation field ($G_{0}$ = 31.6) than derived from the dust SED modelling ($G_{0}$ = 1-3, \citealt{2016MNRAS.459.3900D}). It is plausible that geometry effects play an important role in the determination of the line ratios at the working resolution of a few tens of pc. With the [O~{\sc{i}}] line being detected merely in the very central 9.4$\arcsec$$\times$9.4$\arcsec$ (or 28$\times$28 pc$^{2}$) spaxel of the raster as opposed to the [C~{\sc{ii}}] detection which covers an area of 110 pc $\times$ 75 pc, the actual [O~{\sc{i}}]/[C~{\sc{ii}}] line ratio (accounting for the source sizes) will be lower due to the [O~{\sc{i}}] emission not filling the entire beam. Accounting for the beam filling factors would shift the PDR model parameters ($G_{0}$, $n_{\text{H}}$) towards lower values, compatible with the SED fitting results on spatial scales of 36$\arcsec$ or $\sim$100\,pc. If we include the CO(1-0) line emission that was detected in the high-resolution BIMA CO map presented by \citet{2001AJ....122.1747Y} in the PDR modelling (after correcting the CO(1-0) line emission by a factor of 2 to account for the optically thick CO clouds), we derive PDR model parameters of $G_{0}$ = 1.0 and $n_{\text{H}}$ = 10$^{4.25}$ cm$^{-3}$ which are more consistent with the dust SED modelling results.

The origin of [C~{\sc{ii}}] and [O~{\sc{i}}] emission in NGC\,185 might differ from the classical picture of collisional excitation in PDRs. Turbulent heating by shocks is likely to take place in NGC\,185 based on the observation of shock-excited lines (e.g., H$_{2}$ 0-0 S{0} to S(6), [N~{\sc{ii}}] 6584$\AA$, [S~{\sc{ii}}] 6716,6731$\AA$, [Fe~{\sc{ii}}] 26\,$\mu$m; \citealt{2010ApJ...713..992M,2012MNRAS.419.3159M}). The importance of mechanical heating due to turbulence in shocks has been shown to play an important role in PDRs (e.g., \citealt{2013ApJ...777...66A}), even for low shock velocities \citep{2013A&A...550A.106L}. The low warm-to-cold gas fraction in NGC\,185 (see Section \ref{Warm.sec}) however suggest that the heating through shocks is negligible compared to radiative heating processes. Considering the old nature of the SNR in NGC\,185 with an estimated shock velocity $\leq$ 85 km s$^{-1}$ \citep{2012MNRAS.419..854G}, shock excitation might be able to account for the observed [O~{\sc{i}}] emission in NGC\,185 for shock velocities $\gtrsim$ 35 km s$^{-1}$. The same model could provide at most 10$\%$ of the observed [C~{\sc{ii}}] emission in NGC\,185. In case a significant fraction of the [O~{\sc{i}}] 63\,$\mu$m is excited by shocks, the PDR model parameters would shift to lower $G_{\text{0}}$ and $n_{\text{H}}$. A possible contribution of the old stellar population to the TIR emission, on the other hand, would shift the data points to higher $G_{0}$ and $n_{\text{H}}$ values. 

Based on clear detections of the optical [O~{\sc{i}}] 6300\AA~line \citep{2012MNRAS.419.3159M} with a critical density $n_{\text{crit}}$(H) = 10$^{6}$ cm$^{-3}$ and the infrared [Fe~{\sc{ii}}] 26\,$\mu$m and [Si~{\sc{ii}}] 34.8\,$\mu$m lines \citep{2010ApJ...713..992M} with critical densities of $n_{\text{crit}}$(H) = 2$\times$10$^{6}$ cm$^{-3}$ and $n_{\text{crit}}$(H) = 3$\times$10$^{5}$ cm$^{-3}$, respectively, the presence of a denser ($n_{\text{H}}$ $\sim$ 10$^{4-5}$ cm$^{-3}$) PDR region in the central regions of NGC\,185 is also hinted at. We argue that the filling factor of these dense PDR regions is small compared to the rest of the gaseous ISM with the detection of the [O~{\sc{i}}] 63\,$\mu$m line limited to the central spaxel and based on the small cloud sizes measured from interferometric CO observations \citep{1996ApJ...464L..59Y}. The PDR model parameters derived based on the H$_{2}$ S(0), S(1) and S(2) lines also suggests the presence of gas illuminated by radiation fields $G_{\text{0}}$~$\geq$~3$\times$10$^2$ and gas densities $n_{\text{H}}$ $\sim$ 10$^{4}$ cm$^{-3}$. The detection of several tracers with high excitation temperatures hints at the presence of a gas reservoir exposed to stronger radiation fields or would require alternative excitation mechanisms (e.g., shocks).

To derive a PDR surface temperature (T $\sim$ 40-70 K) for NGC\,185, we rely on the average gas density $n_{\text{H}}$ $\sim$ 10$^{3.75}$ cm$^{-3}$ and a moderate radiation field $G_{0}$ $\sim$1-10 derived from dust SED fitting and PDR modelling. An average PDR temperature of T$\sim$50\,K is used in Section \ref{Cold.sec} to obtain an upper limit for the molecular gas mass traced by [C~{\sc{i}}].

\section{Total gas reservoir}
\label{Totalgas.sec}
We combine the H~{\sc{i}}, CO(1-0), ionised gas, H$_{2}$ and X-ray observations to obtain total gas masses for NGC\,147, NGC\,185 and NGC\,205. Table \ref{ISMmasses} provides an overview of the different gas mass measurements for the three galaxies. 

\begin{table}
\caption{Overview of masses for the different ISM components derived from various observational constraints. The different gas mass measurements are discussed in Section \ref{Totalgas.sec}. The dust masses are taken from \citet{2016MNRAS.459.3900D} for NGC\,147 and NGC\,185 and from \citet{2012MNRAS.423.2359D} for NGC\,205. The total gas mass is obtained by summing the gas masses from different gas phases, scaled by a factor of 1.36 to include Helium, using a Galactic and Z-dependent $X_{\text{CO}}$ conversion factor. The gas-to-dust mass ratios are derived from the global gas and dust mass measurements. The values between parentheses are derived based on $H$-band luminosity-dependent $X_{\text{CO}}$ conversion factors.} 
\label{ISMmasses}
\centering
\begin{tabular}{lccc}
\hline 
Galaxy & NGC\,147 & NGC\,185 & NGC\,205 \\
\hline
$M_{\text{HI}}$ [10$^{4}$ M$_{\odot}$] & $\leq$ 0.4 & 11 & 40 \\
$M_{\text{H}_{2},Gal}$  [10$^{4}$ M$_{\odot}$] & $\leq$ 1.3 & 2.8 &  23 \\
$M_{\text{H}_{2},Z}$  [10$^{4}$ M$_{\odot}$] & $\leq$ 15.8 & 29.1 & 144 \\
$M_{\text{ion}}$ [M$_{\odot}$] &  - & 1 & $\leq$1 \\
warm $M_{\text{H}_{2}}$ [10$^{2}$ M$_{\odot}$] & - & 2.4 & - \\
hot X-ray gas [10$^{4}$ M$_{\odot}$] & $\leq$ 0.2 & $\leq$0.2 &  $\leq$3.8 \\
total $M_{\text{gas,Gal}}$ [10$^{4}$ M$_{\odot}$] & $\leq$2.7 & 18.8 & 85.7  \\  
total $M_{\text{gas,Z}}$ [10$^{4}$ M$_{\odot}$] & $\leq$22.4 & 54.5 & 250.2 \\  
\hline
$M_{\text{dust}}$ [10$^{3}$ M$_{\odot}$] & $\leq$ 0.128 & 5.1 & 11-18 \\
\hline
GDR (Gal. X$_{\text{CO}}$) & - & 37 & 48 \\
GDR (Z-dep. X$_{\text{CO}}$) & - & 107 & 139 \\
\hline 
\end{tabular}
\end{table}

\subsection{H~{\sc{i}} and CO observations}
\label{HIandCO.sec}
For NGC\,147, the 3$\sigma$ upper H~{\sc{i}} mass limit from \citet{1997ApJ...476..127Y} has been scaled to our adopted distance ($M_{\text{HI}}$ $\leq$ 3.7 $\times$ 10$^{3}$ M$_{\odot}$). The 1$\sigma$ upper CO intensity limit ($I_{\text{CO}}$ $\leq$ 0.037 K km s$^{-1}$) reported by \citet{1998ApJ...507..726S} is used to derive a 3$\sigma$ upper limit $M_{\text{H}_{2}}$ $\leq$ 1.3 $\times$ 10$^{4}$ M$_{\odot}$ and $\leq$ 15.8 $\times$ 10$^{4}$ M$_{\odot}$ for a Galactic  ($X_{\text{CO}}$ = 2.0 $\times$ 10$^{20}$ cm$^{-2}$ [K km s$^{-1}$]$^{-1}$) and H-band luminosity-dependent ($X_{\text{CO}}$ = 24.3 $\times$ 10$^{20}$ cm$^{-2}$ [K km s$^{-1}$]$^{-1}$) conversion factor.

For NGC\,185, we adopt the H~{\sc{i}} mass reported in \citet{1997ApJ...476..127Y}, M$_{\text{HI}}$ = 1.1 $\times$ 10$^{5}$ M$_{\odot}$, scaled to our adopted distance. From the CO moment-0 map presented in \citet{2001AJ....122.1747Y}, we derive the H$_{2}$ column density based on N$_{\text{H}_{2}}$ [cm$^{-2}$] = $X_{\text{CO}}$ $I_{CO(1-0)}$ with, $I_{\text{CO(1-0)}}$, the integrated line intensity in units of  K km s$^{-1}$. We apply two types of conversion factors: a Galactic conversion factor ($X_{\text{CO}}$ = 2.0 $\times$ 10$^{20}$ cm$^{-2}$ [K km s$^{-1}$]$^{-1}$, \citealt{1996A&A...308L..21S,2001ApJ...547..792D,2011ApJ...726...81A}) and a $H$-band luminosity-dependent conversion factor ($X_{\text{CO}}$=20.8$\times$10$^{20}$cm$^{-2}$[K km s$^{-1}$]$^{-1}$). Based on these two extremes, we find a molecular gas mass in the range $M_{\text{H}_{2}}$=2.8-29.1$\times$10$^{4}$\,M$_{\odot}$.

In \citet{2012MNRAS.423.2359D}, we derived a total H~{\sc{i}} mass of $M_{\text{HI}}$ $\sim$ 4.0 $\times$ 10$^{5}$ M$_{\odot}$ for NGC\,205. The molecular gas mass ($M_{H_{2}}$ $\sim$ 2.1-13.1 $\times$ 10$^{5}$ M$_{\odot}$) based on CO(1-0) observations of the northern and central regions in NGC\,205 \citep{1996ApJ...464L..59Y,1998ApJ...499..209W} is recalculated for Galactic and $H$-band luminosity-dependent $X_{\text{CO}}$ factors. We, furthermore, add the molecular gas mass derived from the new detections from our NRO 45m observations ($M_{\text{H}_{2}}$ $\sim$ 0.2-1.3 $\times$ 10$^{5}$ M$_{\odot}$, see Section \ref{NRO.sec}), which sums up to a total molecular gas mass of $M_{H_{2}}$ $\sim$ 2.3-14.4 $\times$ 10$^{5}$ M$_{\odot}$.

\subsection{Cold CO-dark molecular gas}
\label{Cold.sec}
Due to the multi-phase origin of the [C~{\sc{ii}}] line emission (see Section \ref{Origin.sec}), we would require a detailed ISM model to disentangle the [C~{\sc{ii}}] emission that is originating from the molecular gas phase. Since such modelling requires an extensive set of multi-phase tracers, we opt to probe the cold molecular gas mass that is not probed by CO observations based on [C~{\sc{i}}] line observations. The \textit{Herschel} SPIRE FTS observations of NGC\,205 allow us to derive 1$\sigma$ upper limits for the undetected [C~{\sc{i}}] line transitions (see Section \ref{SPIREFTS.sec}). Based on those 1$\sigma$ upper limits for the [C~{\sc{i}}] line intensities, we derive upper limits on the molecular gas masses based on Eq. 12 in \citet{2004MNRAS.351..147P}:
\begin{equation}
\label{Eq_CI}
\begin{split}
M_{\text{H}_{2}}~= &~4.92\times10^{10}h'^{-2}\frac{(1+z-\sqrt{1+z})^{2}}{1+z}\left[\frac{X_{\text{CI}}}{10^{-5}}\right]^{-1} \times \\
& \left[\frac{A_{\text{ul}}}{10^{-7}s^{-1}}\right]Q_{\text{ul}}^{-1}\left[\frac{I_{\text{[CI]}}}{\text{Jy}~\text{km}~\text{s}^{-1}}\right]
\end{split}
\end{equation}
where, $h'$, is defined as $H_{\text{0}}$~=~100$h'$ [km s$^{-1}$ Mpc$^{-1}$] with $H_{\text{0}}$=67.8 km s$^{-1}$ Mpc$^{-1}$ \citep{2016A&A...594A..13P}, $z$, is the redshift, $X_{\text{CI}}$, is the neutral carbon abundance, $A_{\text{ul}}$, is the Einstein coefficient for spontaneous emission, $Q_{\text{ul}}$, is the excitation rate coefficient and, $I_{\text{[CI]}}$, is the [C~{\sc{i}}] line intensity. The Einstein coefficients ($A_{\text{10}}$=7.93$\times$10$^{-8}$ s$^{-1}$, $A_{\text{21}}$=2.68$\times$10$^{-7}$ s$^{-1}$, $A_{\text{20}}$=2$\times$10$^{-14}$ s$^{-1}$) are taken from \citet{2004MNRAS.351..147P}. The $Q_{\text{ul}}$ coefficients were calculated under non-local thermal equilibrium (NLTE) conditions for an average gas density $n_{\text{H}}$ $\sim$10$^{3-4}$ cm$^{-3}$ and gas temperature T = 50K (see Section \ref{PDR.sec}) following the recipes from \citet{2004MNRAS.351..147P}, and result in $Q_{\text{10}}$=0.52 and $Q_{\text{21}}$=0.20 for $n_{\text{H}}$=10$^{3}$ cm$^{-3}$, and $Q_{\text{10}}$=0.45 and $Q_{\text{21}}$=0.31 for $n_{\text{H}}$=10$^{4}$ cm$^{-3}$. The [C~{\sc{i}}]/H$_{2}$ abundance ratio is observed to range between values of 10$^{-5}$ and 10$^{-4}$ in local and high-redshift galaxy samples \citep{1989ApJ...344..311F,2002ApJS..139..467I,2003A&A...404..495I,2005A&A...429L..25W,2011ApJ...730...18W}. Simulations show that the neutral carbon abundance mostly varies with metallicity ($X_{\text{[CI]}}$ $\propto$ Z$^{-1}$) and hardly depends on radiation field strength, $G_{\text{0}}$ \citep{2016MNRAS.456.3596G}. We therefore assume a conservative lower limit ([C~{\sc{i}}]/H$_{2}$ $\sim$ 10$^{-5}$) based on the low metallicity of the three dSphs.

Inserting those values in Eq. \ref{Eq_CI}, we derive 1$\sigma$ upper molecular gas mass limits of $M_{\text{H}_{2}}$$\leq$10$^{5}$M$_{\odot}$ for typical gas densities of 10$^{3}$ and 10$^{4}$ cm$^{-3}$ in the CO peak in the north of NGC\,205. Similar upper molecular gas mass limits of $M_{\text{H}_{2}}$$\leq$10$^{5}$M$_{\odot}$ are derived in the centre of NGC\,205. We only provide the gas masses derived from the [C~{\sc{i}}] 1-0 line transitions, since they give the tightest upper limits. These 1$\sigma$ upper mass limits are smaller than the molecular gas masses estimated from PDR models. For gas densities of $n_{\text{H}}$ $\sim$ 10$^{3-4}$ cm$^{-3}$ and a radiation field $G_{\text{0}}$ $\in$ [1,10], the PDR models from \citet{1999ApJ...527..795K} predict line intensity ratios of [C~{\sc{ii}}]/[C~{\sc{i}}]~1-0 $\sim$ 1-10 and [C~{\sc{i}}]~2-1/[C~{\sc{i}}]~1-0 $\sim$ 2 (see plots in the PDR toolbox, \citealt{2008ASPC..394..654P}). We derive a [C~{\sc{ii}}] 158\,$\mu$m line flux of 1.4$\times$10$^{-16}$ W m$^{-2}$ (or 2212 Jy km s$^{-1}$) for NGC\,205 from \textit{Herschel} observations \citep{2012MNRAS.423.2359D}. PDR models would thus predict line intensities of [C~{\sc{i}}] 1-0 $\sim$ 221-2212 Jy km s$^{-1}$ and [C~{\sc{i}}] 2-1 $\sim$ 110-1106 Jy km s$^{-1}$. Based on the PDR model predictions, we would expect both [C~{\sc{i}}] line transitions to be detected. The non-detection of the neutral carbon line transitions in NGC\,205 might reflect a difference in filling factors between the ISM phases probed by [C~{\sc{ii}}] and [C~{\sc{i}}]. Studies of low-metallicity star-forming dwarf galaxies show that the filling factor of the cold dense molecular gas phase probed by [C~{\sc{i}}] is lower compared to the ISM phases probed by [C~{\sc{ii}}] (De Looze et al.\,in prep.), which might also explain the non-detection of the [C~{\sc{i}}] lines in NGC\,205. With the lowest CO transitions probing the cold and dense molecular gas, the filling factor of [C~{\sc{i}}]-emitting clouds is thought to be more compatible with the CO-emitting surfaces of dense clouds. Based on the integrated CO 1-0 line intensity (0.39 K km s$^{-1}$) observed with the NRAO 12m telescope (FWHM=55$\arcsec$) in the centre of NGC\,205 \citep{1998ApJ...499..209W} and the theoretical line ratio [C~{\sc{i}}] 1-0/CO 1-0 $\sim$ 25 derived for a PDR model with (G$_{\text{0}}$, n$_{\text{H}}$) = (1, 10$^{3.75}$ cm$^{-3}$), we can put a limit on the line intensity I$_{\text{[CI] 1-0}}$ $\sim$ 10$^{-18}$ W m$^{-2}$ in the centre of NGC\,205 which is comparable to the 1$\sigma$ noise level in the SPIRE FTS spectra and consistent with the non-detection of [C~{\sc{i}}] in the smaller SPIRE FTS beam (FWHM=40$\arcsec$). Earlier works have already shown that the intensity and/or cloud filling factor of [C~{\sc{i}}] line emission predicted by PDR models can be inconsistent with observations and suggest a smaller filling factor for denser clouds (e.g., \citealt{1985ApJ...299..967K,2014ApJ...781..101S}). 

We can furthermore derive an upper limit on the CO-dark molecular gas reservoir based on the \textit{Herschel} [C~{\sc{ii}}] observations of NGC\,185. If we account for the negligible [C~{\sc{ii}}] contribution from ionised gas and the estimated [C~{\sc{ii}}] emission from neutral atomic gas (see Section \ref{Origin.sec}), we can invert Eq. \ref{EqMadden} to derive the C$^{+}$ column density, $N_{\text{C+}}$, in the molecular gas phase from the residual [C~{\sc{ii}}] emission after subtracting the [C~{\sc{ii}}] contribution from neutral atomic gas clouds. When adopting a critical density of $n_{\text{crit}}$\,=\,7600 cm$^{-3}$ for collisions with H$_{2}$ molecules (see \citealt{2012ApJS..203...13G}, derived for a kinetic gas temperature T=20\,K typical of molecular clouds) and assuming the best-fitting PDR model parameters ($n_{\text{H}}$=10$^{3.75}$ cm$^{-3}$ and T=50\,K, see Section \ref{PDR.sec}), we derive a molecular gas mass of $\sim$4$\times$10$^{4}$ M$_{\odot}$ for the [C~{\sc{ii}}] emitting gas. Assuming a lower PDR gas density (10$^3$ cm$^{-3}$) could increase this molecular gas mass by a factor of $\sim$4. The latter value should be regarded a strict upper limit since part of the [C~{\sc{ii}}] emission might originate from neutral atomic gas clouds rather than the molecular gas phase in PDRs. The latter upper limits on the CO-dark molecular gas reservoir are consistent with the upper limit of $\leq$10$^{5}$M$_{\odot}$ derived based on the non-detection of [C~{\sc{i}}] line emission in NGC\,205.

The [C~{\sc{i}}]-based upper limits of the molecular gas mass ($M_{\text{H}_{2}}$$\leq$1$\times$10$^{5}$M$_{\odot}$) in NGC\,205 are smaller compared to the H$_{\text{2}}$ masses probed by CO ($M_{\text{H}_{2}}$=2.8-29.1$\times$10$^{5}$M$_{\odot}$). We thus conclude that the CO-dark molecular gas fraction is small in NGC\,205 and that the CO line emission traces the bulk of molecular gas. Given that the ISM conditions are very similar in NGC\,185, we believe that the latter argument can also be applied to NGC\,185.

\subsection{Warm molecular gas}
\label{Warm.sec}
The detection of rotational transitions of molecular hydrogen with the \textit{Spitzer} IRS spectrometer \citep{2010ApJ...713..992M} suggests the presence of a reservoir of warm molecular gas in the centre of NGC\,185, significantly warmer than the PDR component traced by [C~{\sc{ii}}] and [O~{\sc{i}}]. Based on the observed intensities of rotational H$_{2}$ transitions for the three \textit{Spitzer} IRS pointings in NGC\,185 (see Fig. \ref{Ima_NGC185_AOR}) reported by \citet{2010ApJ...713..992M}, we infer the temperature and column density of the warm molecular gas phase. Different mechanisms are capable of exciting H$_{2}$ molecules, among which radiative excitation by massive stars with photon energies 6 $\leq$ h$\nu$ $\leq$ 13.6 eV in PDRs, and shock excitation in molecular outflows and supernova remnants are the most important contributors. 

The detection of rotational H$_{2}$ transitions up to S(7) (with excitation temperatures up to 5828 K) seems unlikely to be driven by strong radiation (given the soft radiation field $G_{\text{0}}$ $\sim$ 1-3 derived from dust SED modelling, \citealt{2016MNRAS.459.3900D}) and suggests that the highest H$_{2}$ transitions are mainly shock excited. We, therefore, restrict the fitting of representative temperatures and column densities to the lower rotational levels of H$_{2}$ (S(0) to S(2)) which are generally in collisional equilibrium \citep{1992ApJ...399..563B}.  

\begin{figure*}
\centering
\includegraphics[width=5.7cm]{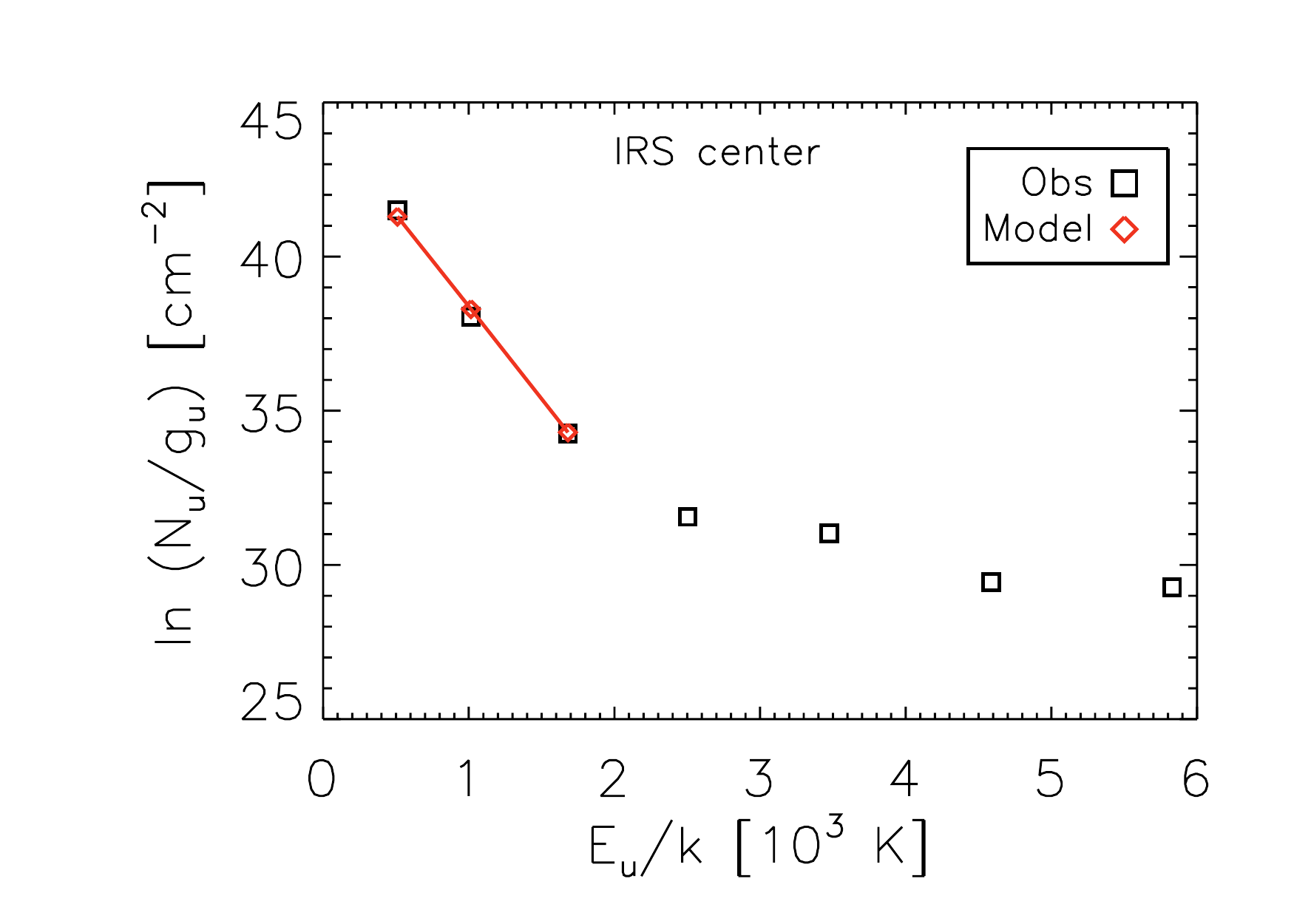}   
\includegraphics[width=5.7cm]{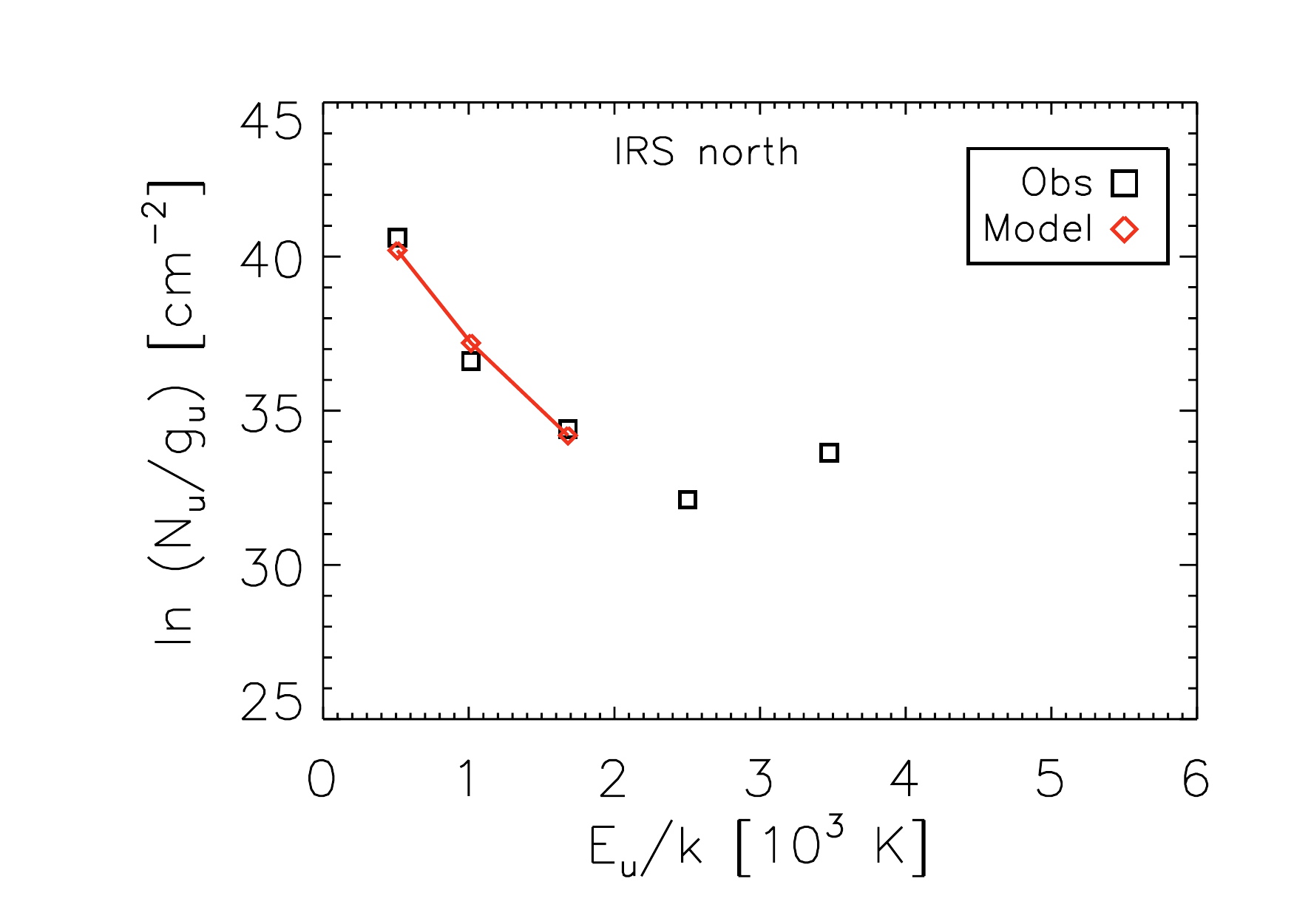}   
\includegraphics[width=5.7cm]{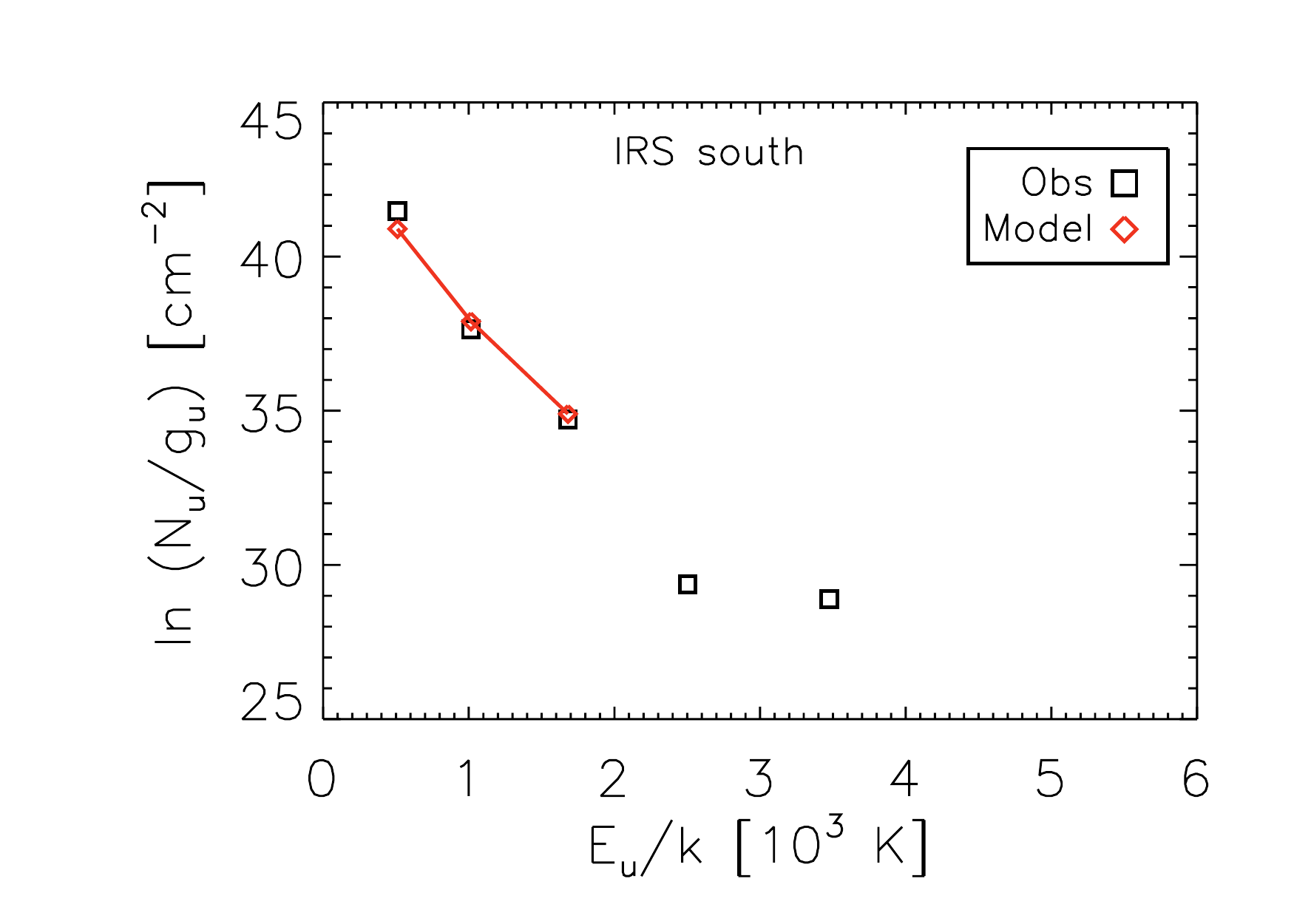}   \\
\includegraphics[width=5.7cm]{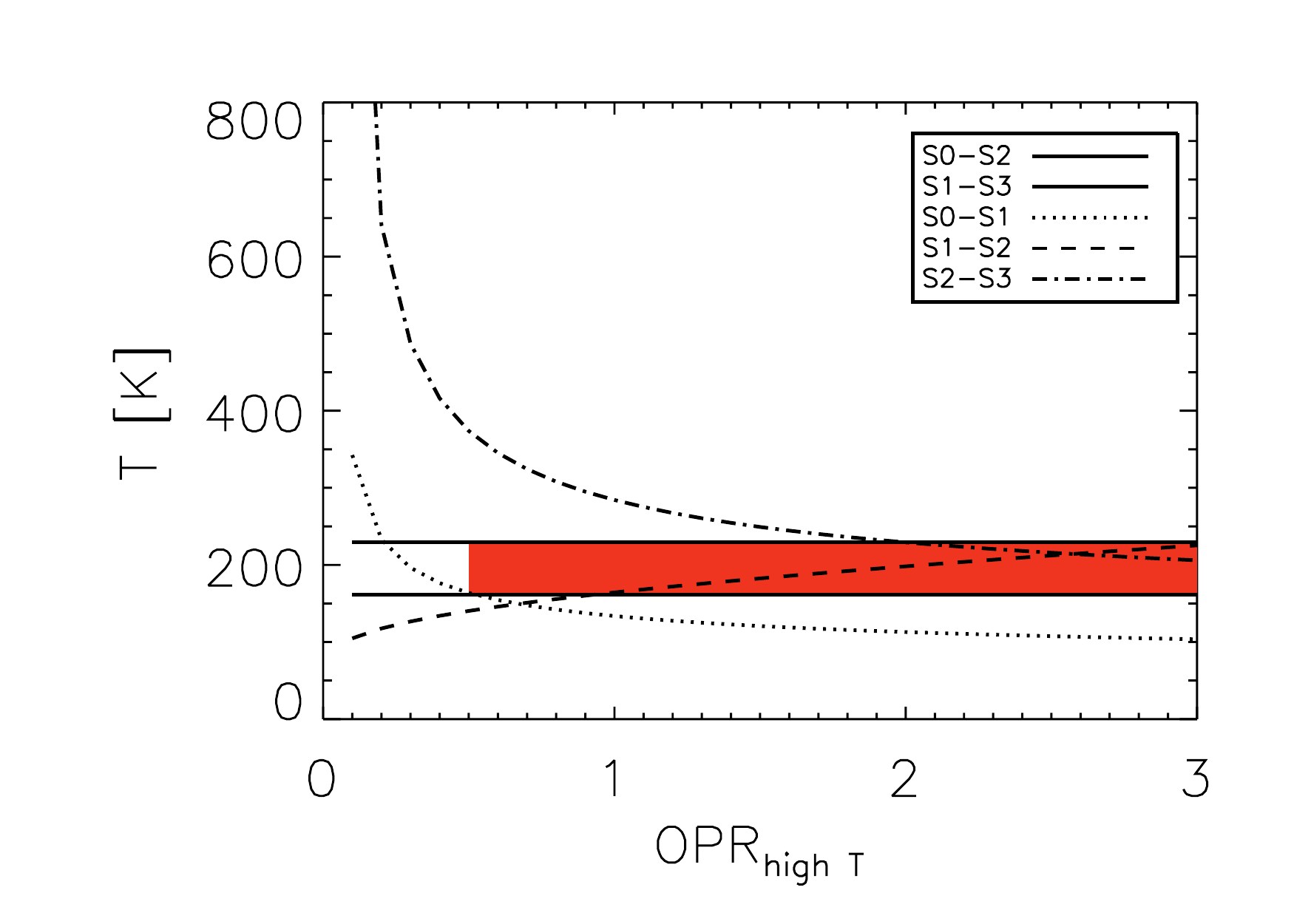}   
\includegraphics[width=5.7cm]{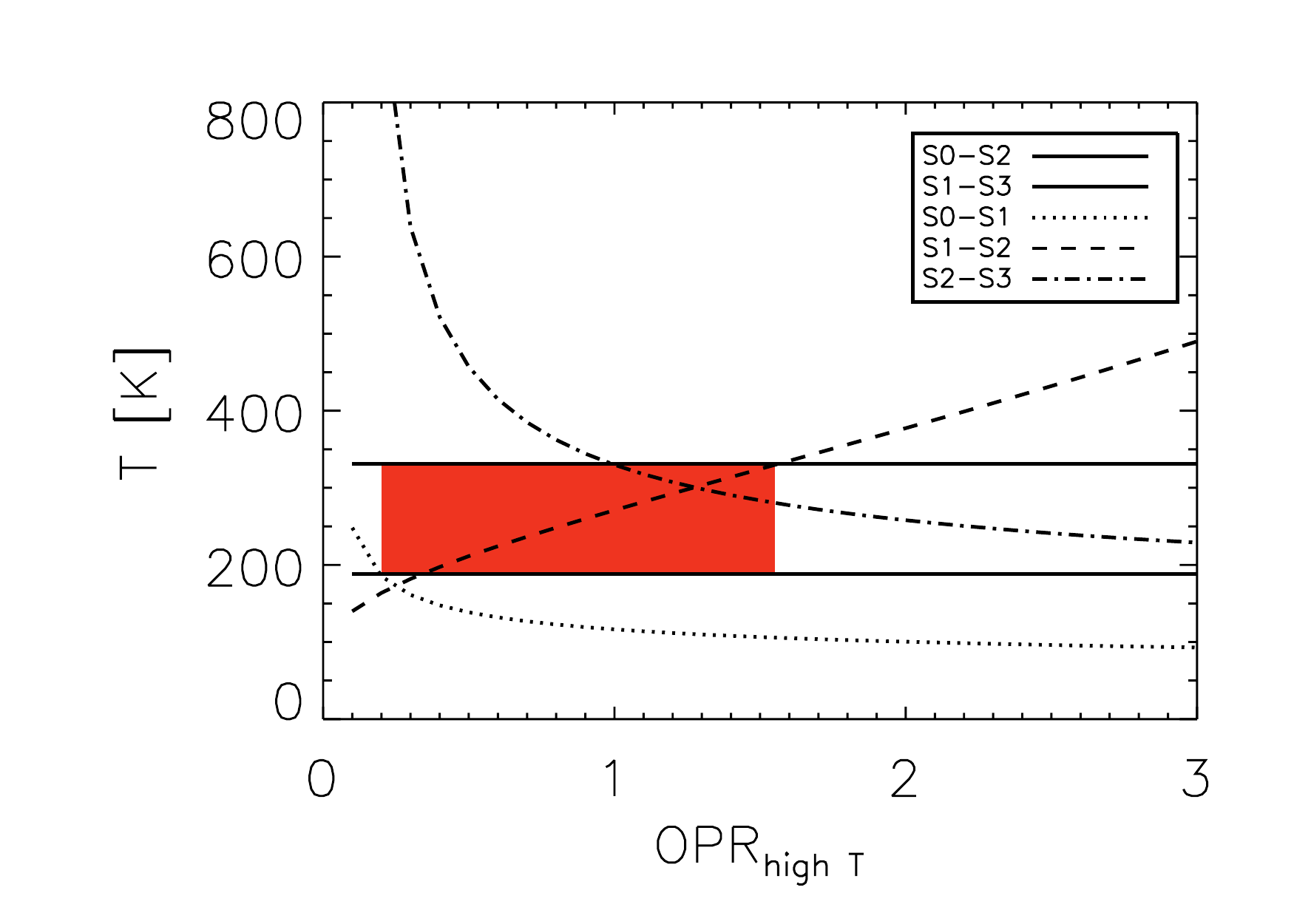}   
\includegraphics[width=5.7cm]{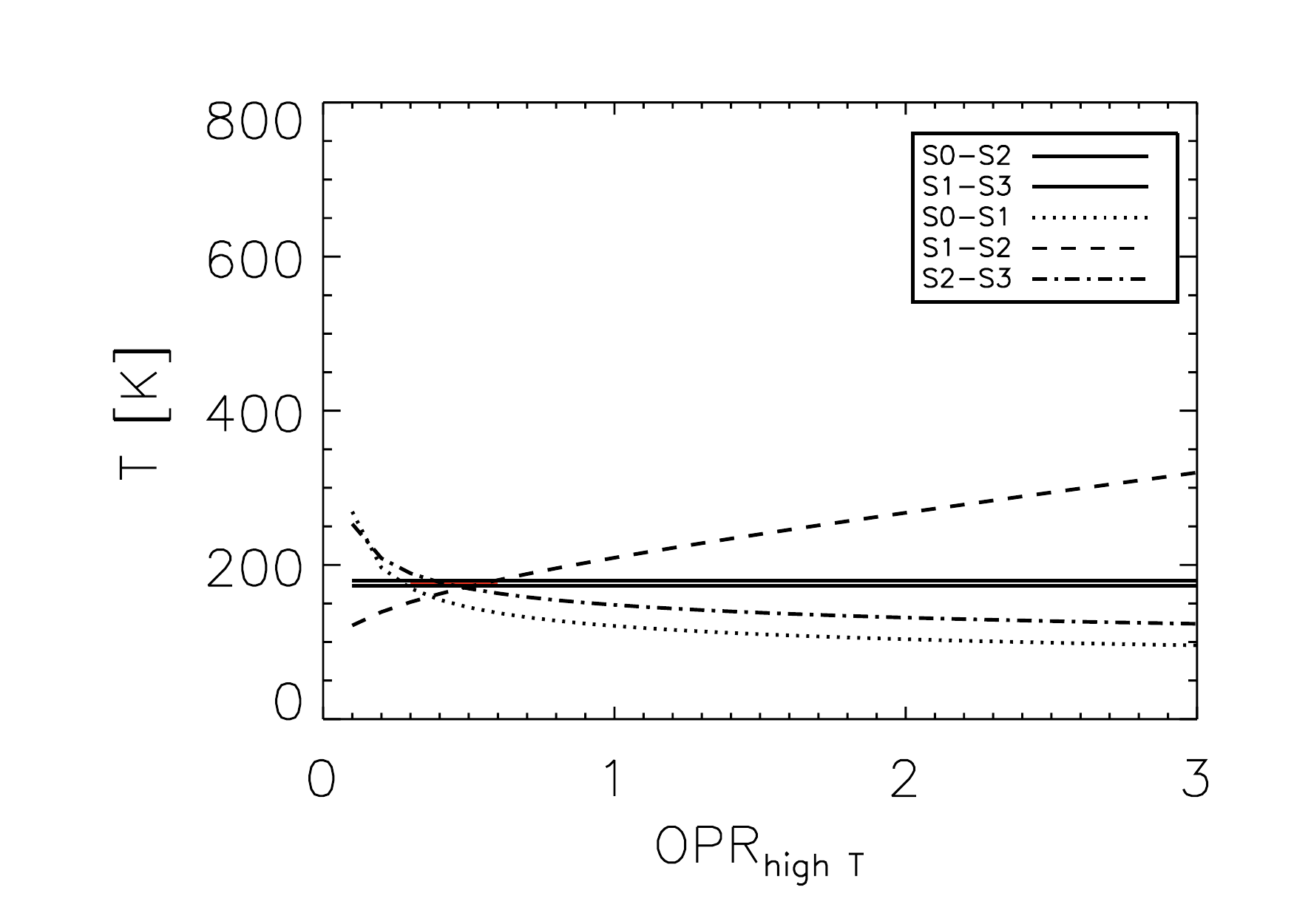}    
\caption{Top: Observed H$_{2}$ excitation diagrams in the central (left), north (middle), and south (right) IRS pointings for NGC\,185. The red line and symbols indicate the best fitting model with excitation temperature, $T_{\text{ex}}$, and column density, N$_{\text{H}_{2}}$, fitted to the first three H$_{2}$ rotational transitions S(0), S(1), and S(2)), for an ortho-to-para ratio of 3. Bottom: Diagram of the excitation temperatures derived for different pairs of rotational H$_{2}$ transitions. The excitation temperatures should increase monotonically with higher order transitions to be compatible with the thermalization of H$_{2}$ lines. }
             \label{ima_excdiag}
\end{figure*}

Figure \ref{ima_excdiag} (top row) presents the excitation diagrams for the central, north and south region in NGC\,185, respectively. Excitation diagrams visualise the distribution of different level populations described by the column density of the upper state $N_{u}$ divided by the statistical weight $g_{u}$ of that level population as a function of the upper state energy level $E_{u}$/$k$. For the construction of this excitation diagram, we have assumed that optical depth effects are negligible, which should be appropriate given the low metal abundance of NGC\,185. 

The best fit to the S(0), S(1) and S(2) lines results in a temperature $T$ $\sim$ 180 K and column density $N_{\text{H}_{2}}$ $\sim$ 2.0 $\times$ 10$^{18}$ cm$^{-2}$ in the central region of NGC\,185. Within the \textit{Spitzer} IRS extraction area (35.7$\arcsec$ $\times$ 10.7$\arcsec$), the column density corresponds to a mass $M_{\text{H}_{2}}$ $\sim$ 111 M$_{\odot}$ of warm molecular gas. This warm molecular gas mass should be regarded as a lower limit, since the area covered by the IRS slit is limited. A similar analysis of the lower $H_{2}$ transitions in the excitation diagram for the IRS slit positions observed in the north and south of the galaxy indicates warm molecular gas masses of $M_{\text{H}_{2}}$ $\sim$ 44 M$_{\odot}$ (for best fitting parameters $T$ $\sim$ 190 K and $N_{\text{H}_{2}}$ $\sim$ 8.0 $\times$ 10$^{17}$ cm$^{-2}$) in the northern pointing and $M_{\text{H}_{2}}$ $\sim$ 87 M$_{\odot}$ (for best fitting parameters $T$ $\sim$ 190 K and $N_{\text{H}_{2}}$ $\sim$ 1.6 $\times$ 10$^{18}$ cm$^{-2}$) towards the south of NGC\,185. The temperatures derived for the different regions in NGC\,185 are lower compared to the average temperatures ($T$ $\sim$ 350-380 K) derived for low-metallicity star-forming dwarf galaxies in \citet{2014A&A...564A.121C} due to the harder and stronger radiation fields in those star-forming dwarfs, but similar to the cold molecular gas temperatures (T $\sim$ 150 K) derived for normal spiral galaxies \citep{2007ApJ...669..959R}. 

For the determination of the best-fitting temperature and column density from the observed excitation diagram, we assumed that the condition of local thermal equilibrium (LTE) is fulfilled for the lowest H$_{2}$ transitions. Under LTE conditions, we expect to derive lower excitation temperatures for ratios of transitions with lower energy upper levels for an ortho-to-para density ratio of 3 \citep{1992ApJ...399..563B} or, explicitly, T(S(1)-S(2)) $<$ T(S(1)-S(3)) $<$ T(S(2)-S(3)). Following the procedure in \citet{2007ApJ...669..959R}, we can determine the excitation temperature of consecutive transitions as a function of the ortho-to-para ratio (OPR) and, hereby, verify whether the diagram shows departures from thermalisation of ortho-to-para levels.  Figure \ref{ima_excdiag} (bottom panels) shows the determined excitation temperatures as a function of OPR for each pair of transitions from S(0) to S(3), for the central, north and south IRS positions in NGC\,185. The red-colored region satisfies the thermalisation condition. The thermalisation of H$_{2}$ transitions up to S(3) seems satisfied only for the central region. Given that the S(3) - S(2) ratio is not consistent with OPR = 3 for the northern and southern regions, the higher rotational transition of $H_{2}$ in these cases no longer satisfies collisional equilibrium and is likely excited by shocks. 

Given the violation of the LTE conditions in the north and south IRS positions of NGC\,185, we might overestimate the excitation temperature (due to a shock contribution) and underestimate the warm molecular gas mass in those regions. Even though the entire volume of warm molecular gas could not be traced due to the limited \textit{Spitzer} IRS coverage, the observed warm-to-cold molecular gas fractions (ranging from 0.001 to 0.01) are one to two orders of magnitude lower than the typical warm-to-cold gas fractions reported by \citet{2007ApJ...669..959R}. We, therefore, do not expect to find massive reservoirs of warm molecular gas in NGC\,185.

Due to the lack of IRS spectra for NGC\,147 and NGC\,205, we are not able to put constraints on the warm molecular gas reservoir in those galaxies.

\subsection{Ionised gas}
\label{ion.sec}
We estimate the ionised hydrogen mass from the observed H$\alpha$ luminosity ($L_{\text{H}\alpha}$ $\sim$ 1.3 $\times$ 10$^{36}$ erg s$^{-1}$, \citealt{1999AJ....118.2229M}) in NGC\,185 and convert it into an H~{\sc{ii}} mass (see Eq. 5 from \citealt{2010MNRAS.407.2475F}) assuming an electron temperature $T_{\text{e}}$ = 10$^{4}$ K and electron density $N_{\text{e}}$ $\sim$ 8,300 cm$^{-3}$ (which is the mean electron density determined for three PNe in NGC\,185 by \citealt{2012MNRAS.419..854G}). The resulting H~{\sc{ii}} mass $M_{\text{H~II}}$ = 1 M$_{\odot}$ is negligible compared to the neutral gas mass in NGC\,185.

Similarly, the 1$\sigma$ noise level (2.5$\times$10$^{-17}$ erg s$^{-1}$ cm$^{-2}$ arcsec$^{-2}$) in the H$\alpha$ image of NGC\,205 \citep{1997ApJ...476..127Y} can be converted into a 3$\sigma$ upper mass limit $M_{\text{H~II}}$ $\leq$ 1 M$_{\odot}$ assuming an electron temperature $T_{\text{e}}$ = 10$^{4}$ K and electron density $N_{\text{e}}$ $\sim$ 5,300 cm$^{-3}$ (which is the mean electron density of PNe determined by \citealt{2014MNRAS.444.1705G}).

We are not aware of any H$\alpha$ observations for NGC\,147 which would allow us to put a constraint on the ionised gas mass in this galaxy.

\subsection{Hot X-ray gas}
\label{X-ray.sec}
The non-detections of X-ray emission from NGC\,147 and NGC\,185 \citep{1997MNRAS.291..709B}, and NGC\,205 \citep{1998ApJ...499..209W} allow us to put a constraint on the reservoir of hot gas in these dwarf spheroidal galaxies. We calculate the upper X-ray gas mass limit based on the prescriptions from \citet{1991ApJS...75..751R} and the X-ray and B-band\footnote{B-band luminosities are determined from the RC3 flux densities reported on NED.} luminosities of the galaxies. We find 1$\sigma$ upper limits on the X-ray gas mass of $\leq$ 0.2, 0.2, 3.8 $\times$ 10$^{4}$ for NGC\,147, NGC\,185, and NGC\,205, respectively. 

\subsection{Total gas mass}
For the computation of the total gas mass, we combine H~{\sc{i}} gas, CO-traced and CO-dark molecular gas, ionised and hot X-ray gas masses, and the warm molecular gas reservoir cooled by H$_{2}$ rotational lines (see Table \ref{ISMmasses}). The latter gas reservoirs combine to total gas masses of $M_{\text{g}}$ = 1.9$\times$10$^{5}$\,M$_{\odot}$ and 8.6$\times$10$^{5}$ M$_{\odot}$ (corrected by a factor of 1.36 to account for helium) for NGC\,185 and NGC\,205, respectively, and an upper limit of $M_{\text{g}}$ $\leq$ 2.7$\times$10$^{4}$\,M$_{\odot}$ for NGC\,147, assuming Galactic $X_{\text{CO}}$ conversion factors. Using metallicity-dependent $X_{\text{CO}}$ factors, the total gas masses are up to three times higher with gas masses of $M_{\text{g}}$ = 5.5$\times$10$^{5}$\,M$_{\odot}$ and $M_{\text{g}}$ = 25.0$\times$10$^{5}$\,M$_{\odot}$ for NGC\,185 and NGC\,205, and an upper limit of $M_{\text{g}}$ $\leq$ 2.7$\times$10$^{4}$\,M$_{\odot}$ for NGC\,147. If we assume Galactic $X_{\text{CO}}$ factors, the gaseous reservoirs in NGC\,185 and NGC\,205 are dominated by the atomic H~{\sc{i}} gas component, while a metallicity-dependent $X_{\text{CO}}$ factor would imply molecular gas reservoirs that are 3 to 4 times more massive than the atomic hydrogen content.


\section{Star formation efficiency}
\label{SFE.sec}
Based on the gas mass measurements from Section \ref{Totalgas.sec}, we can link the reservoir that is available for star formation to the actual star formation rate, to learn more about the efficiency of gas consumption in dwarf spheroidal galaxies. We derive total and molecular gas depletion time scales $\tau_{dep} = \Sigma_{gas}/\Sigma_{SFR}$, which is the time needed to exhaust the current total (H~{\sc{i}}+H$_{\text{2}}$) and molecular (H$_{\text{2}}$) gas reservoir. Assuming that most of the gas content is located in the central regions (150 pc $\times$ 90 pc) where recent star formation (6.6 $\times$ 10$^{-4}$ M$_{\odot}$ yr$^{-1}$) took place, we derive gas depletion time scales of $\tau_{\text{HI+H}_{2}}$ $\sim$ 0.3\,Gyr (0.8\,Gyr) and $\tau_{\text{H}_{2}}$ $\sim$ 0.06 Gyr (0.6 Gyr) for a Galactic ($H$ band luminosity-dependent) $X_{\text{CO}}$ factor for NGC\,185. Due to the lack of any recent star formation activity and the non-detection of any gas in NGC\,147, it is impossible to calculate a gas depletion time scale for this galaxy.

We use constraints on the SFR $\sim$ 7 $\times$ 10$^{-4}$ M$_{\odot}$ yr$^{-1}$ \citep{2009A&A...502L...9M} in the central 28$\arcsec$ $\times$ 26$\arcsec$ region of NGC\,205. We calculate the total gas mass (2.0 $\times$ 10$^{5}$ M$_{\odot}$) in this central region of NGC\,205 based on the atomic hydrogen mass ($M_{\text{HI}}$ $\sim$ 7.6 $\times$ 10$^{4}$ M$_{\odot}$) of two central H~{\sc{i}} clumps \citep{1997ApJ...476..127Y} and the molecular gas mass ($M_{\text{H}_{2}}$ $\sim$ 6.8 $\times$ 10$^{4}$ M$_{\odot}$, assuming a Galactic $X_{\text{CO}}$ factor) from the central CO pointing of \citet{1998ApJ...499..209W}, scaled by a factor of 1.36 to include helium. Using a $H$-band luminosity-dependent $X_{\text{CO}}$ factor (12.5 $\times$ 10$^{20}$ cm$^{-2}$ [K km s$^{-1}$]$^{-1}$) would imply a molecular gas reservoir of 4.2 $\times$ 10$^{5}$ M$_{\odot}$. We derive gas depletion time scales of $\tau_{\text{HI+H}_{2}}$ $\sim$ 0.3\,Gyr (1.0\,Gyr ) and $\tau_{\text{H}_{2}}$ $\sim$ 0.1\,Gyr (0.8\,Gyr) for a Galactic ($H$ band luminosity-dependent) $X_{\text{CO}}$ factor for NGC\,205.

\begin{figure}
\centering
\includegraphics[width=8.5cm]{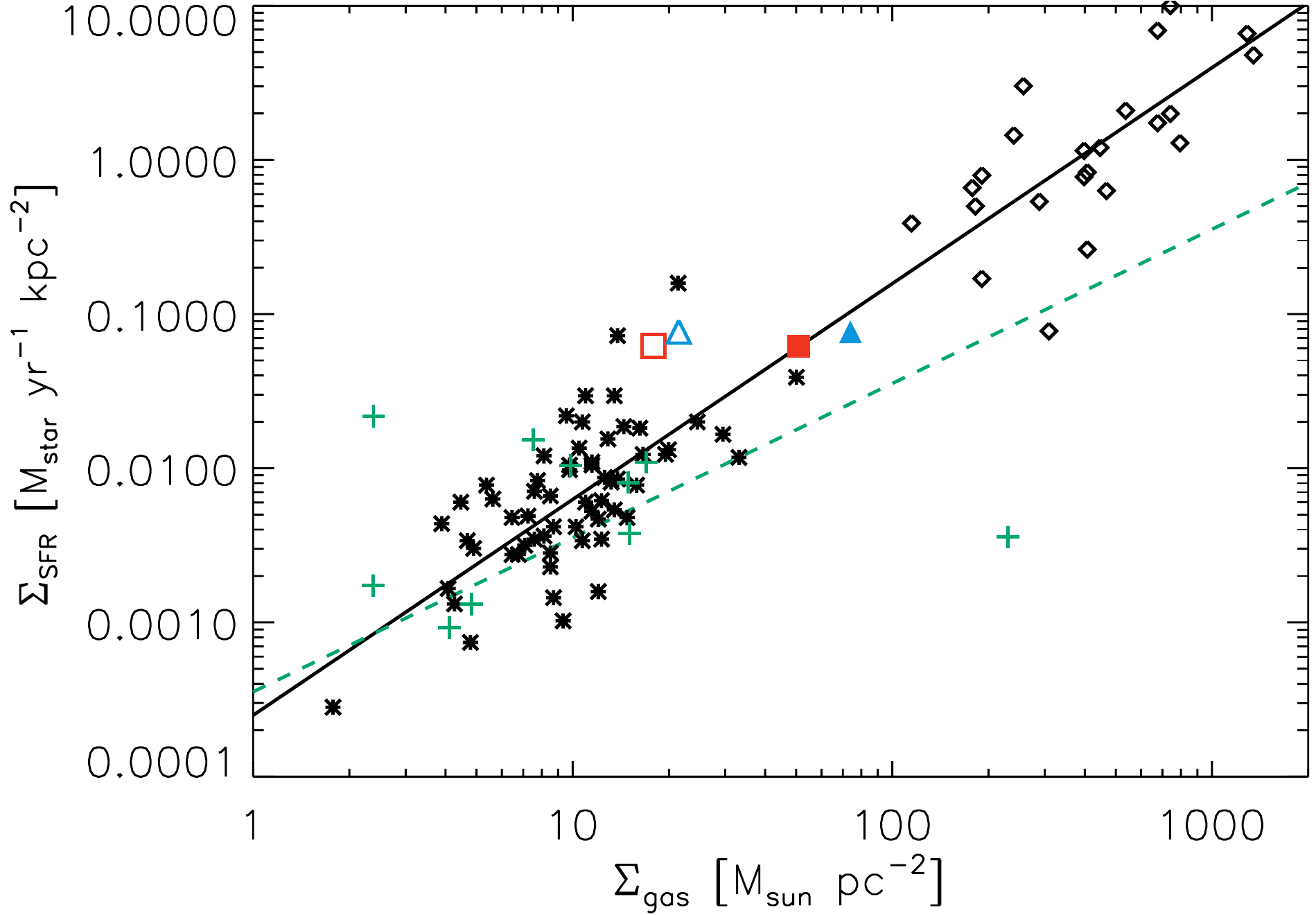}   \\
\caption{The Kennicutt-Schmidt relation between the gas surface density and the SFR surface density with a representative sample of spiral and starburst galaxies from \citet{1998ApJ...498..541K} indicated as black asterisks and diamonds, respectively. The locations of metal-poor star-forming dwarf galaxies studied by \citet{2014A&A...564A.121C} are indicated with green crosses in the K-S diagram (we omitted two galaxies with highly uncertain molecular gas masses). We only included the dwarfs from \citet{2014A&A...564A.121C} with constraints on H~{\sc{i}} and H$_{2}$ masses, and scaling CO intensities with metallicity-dependent $X_{\text{CO,Z}}$ factors to obtain the latter masses. The dwarf spheroidal galaxies under analysis in this paper, NGC\,185 and NGC\,205, are marked as red squares and blue triangles. Empty symbols represent total gas masses (H~{\sc{i}}+H$_{2}$) with Galactic $X_{\text{CO}}$ conversion factors, while filled symbols indicate total gas surface densities calculated with $H$ band luminosity dependent $X_{\text{CO}}$ factors. The black solid line represents the K-S relation $\Sigma_{SFR}$ = A $\Sigma_{gas}^{N}$ with $N = 1.4$ as found by \citet{1998ApJ...498..541K} where $\Sigma_{gas}$ is the total gas (H~{\sc{i}}+H$_{2}$) surface density, while the green dashed line represents the average molecular gas depletion time scale $\sim$ 2 Gyr in a sample of nearby spiral galaxies derived by \citet{2013AJ....146...19L}.}
            \label{Ima_KSlaw}
\end{figure}

Based on these gas mass depletion time scales, the dwarf spheroidal galaxies, NGC\,185 and NGC\,205, are forming stars more actively in comparison to galaxies on the \citet{1998ApJ...498..541K} relation. At the current depletion time scale, the entire molecular gas reservoir (based on a Galactic $X_{\text{CO}}$ factor) would be exhausted within less than 100 Myr. For metallicity-dependent $X_{\text{CO}}$ factors, the star formation efficiencies are more comparable to normal spiral galaxies, but molecular gas depletion times scales still about a factor of 2 lower compared to the average $\Sigma_{\text{H}_{2}}$ $\sim$ 2 Gyr in a sample of nearby spiral galaxies derived by \citet{2013AJ....146...19L}.
Although the level of star formation in NGC\,185 (SFR $\sim$ 6.6 $\times$ 10$^{-4}$ M$_{\odot}$) and NGC\,205 (SFR $\sim$ 7 $\times$ 10$^{-4}$ M$_{\odot}$) is very different from the star formation rates observed in normal star-forming galaxies (1-10 M$_{\odot}$ yr$^{-1}$), the SFR conditions in the centres of those dSphs do seem to approach the star-forming conditions observed in local spirals (and might be even more efficient). Independent evidence for a high star formation efficiency in NGC\,185 was derived from its observed abundance [O/Fe] ratio (0.8 dex, \citealt{2012MNRAS.419..854G}), which indicates a higher SNe II rate compared to SNe Ia. The positive [O/Fe] ratio suggests that most of the gas has been consumed on short time scales. 

Alternatively, we might be observing these two dSph galaxies towards the end of their recent star formation episode, having already burned most of their initial gas reservoir and resulting in an artificially high SFE estimate. We also caution that the SFE predictions might be affected by stochastic effects due to the small number of clouds detected within these galaxies.

\section{Gas-to-dust mass ratio}
\label{GDR.sec}
Combining all gas and dust mass measurements (see Table \ref{ISMmasses}), we derive estimates of the global gas-to-dust mass ratios GDR$\sim$37 in NGC\,185 and GDR$\sim$48 in NGC\,205. Using metallicity-dependent $X_{\text{CO}}$ factors, the global gas-to-dust mass ratio would increase to GDR$\sim$107 for NGC\,185 and GDR$\sim$139 for NGC\,205. Since the H~{\sc{i}} gas is more extended compared to the dust in NGC\,185 and NGC\,205 (see Figures 2 and 4, respectively), the gas-to-dust mass ratio might become even smaller on local scales. These global values are lower than the average Galactic gas-to-dust mass ratio $\sim$ 130 \citep{2007ApJ...657..810D}. Based on the observed trend of increased gas-to-dust mass ratios with decreasing metallicity (e.g., \citealt{1998ApJ...496..145L,2002MNRAS.335..753J,2005A&A...434..849H,2008ApJ...678..804E,2008ApJ...672..214G,2011A&A...532A..56G,2011A&A...535A..13M,2014A&A...563A..31R}), these low gas-to-dust mass ratios are considered even more exceptional, where a simple GDR $\propto$ Z$^{-1}$ scaling would imply a GDR$\sim$370 for NGC\,185 and GDR$\sim$520 for NGC\,205. 

Similarly low gas-to-dust mass ratios were observed in the elliptical galaxy, NGC\,4125 \citep{2013ApJ...776L..30W}, and the dust-lane lenticular galaxy, NGC\,5485 \citep{2014MNRAS.444L..90B}. The low gas-to-dust mass ratio in NGC\,4125 was attributed to the rapid heating of gas to temperatures $\gtrsim$ 10$^{4}$ K, faster than the evaporation of cold dust in this galaxy \citep{2013ApJ...776L..30W}. Such a scenario seems, however, unlikely for NGC\,185, where the warm-to-cold molecular gas fractions (0.001-0.01) are much lower than observed in more massive star-forming galaxies. Also the non-detection of X-ray emission (see Section \ref{X-ray.sec}) in NGC\,185 and NGC\,205 is able to put an upper limit on the reservoir of hot gas ($\lesssim$ 0.2-3.8 $\times$ 10$^{4}$ M$_{\odot}$). 

None of the chemical evolution models, including interstellar grain growth \citep{2013EP&S...65..213A} and accounting for a wide variation of star formation histories \citep{2014A&A...562A..76Z} (see Figures 8 and 9 in \citealt{2014A&A...563A..31R}), predict a gas-to-dust ratio as low as that observed in these dSphs considering its metal abundance. Since these low GDRs clearly deviate from theoretical model predictions, it is worth investigating the origin of the discrepancy between model and observations. 

First, we consider possible caveats in the determination of total gas and dust masses. Given that the H~{\sc{i}} and CO observations are sufficiently deep to detect faint emission \citep{1997ApJ...476..127Y,2001AJ....122.1747Y}, we are confident that the current H~{\sc{i}} and CO data sets will not miss a massive reservoir of atomic or molecular gas. The warm molecular gas masses might be underestimated due to model assumptions and/or insufficient observational coverage. It is, however, unrealistic to assume that the H$_{2}$ observations with \textit{Spitzer} can account for a substantial massive gas reservoir given the low warm-to-cold molecular ratio (see Section \ref{Warm.sec}). The presence of a massive ionised gas reservoir is also unlikely given the weak H$\alpha$ emission from NGC\,185. The non-detection of [C~{\sc{i}}] in NGC\,205 implies that the CO-dark molecular gas content is insignificant compared to the H$_{\text{2}}$ mass traced by CO.

Although the dust masses in NGC\,185 and NGC\,205 have been robustly measured in \citet{2016MNRAS.459.3900D} and \citet{2012MNRAS.423.2359D}, the lack of knowledge on the dust composition and dust mass absorption coefficients makes the derived dust masses uncertain by at least a factor of 2. Even with this uncertainty factor of two, the main cause for the low gas-to-dust mass ratios seems hard to explain based on a lack of observational constraints and/or inaccuracies in the ISM mass predictions. We attribute the low gas-to-dust mass ratios to a combination of possible effects including efficient dust production and longterm grain survival (see \citealt{2016MNRAS.459.3900D} and Section \ref{GasSource.sec}), and the removal of part of the gas mass from the galaxy (see Section \ref{Tidal.sec}). 

\section{The ISM mass budget}
\label{MissingMass.sec}
In this section, we discuss the origin of gas and dust reservoirs in NGC\,147, NGC\,185 and NGC\,205. Hereto, we compare the gaseous reservoirs detected in these galaxies to theoretical predictions from a simple closed-box model. 

\subsection{Theoretical gas consumption and replenishment}
\label{GasSource.sec}

\begin{figure*}
\centering
\includegraphics[width=8.0cm]{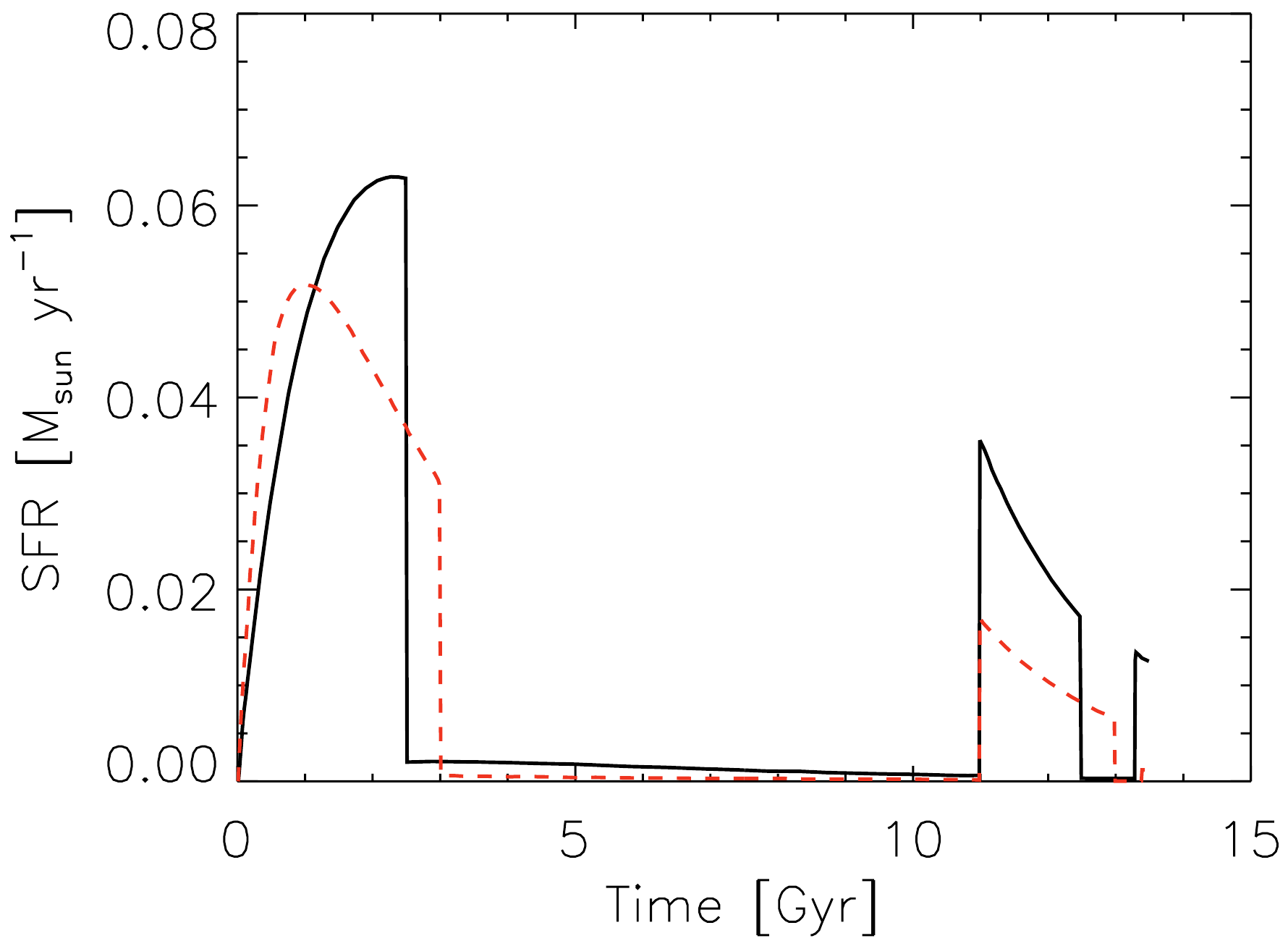}   
\includegraphics[width=7.7cm]{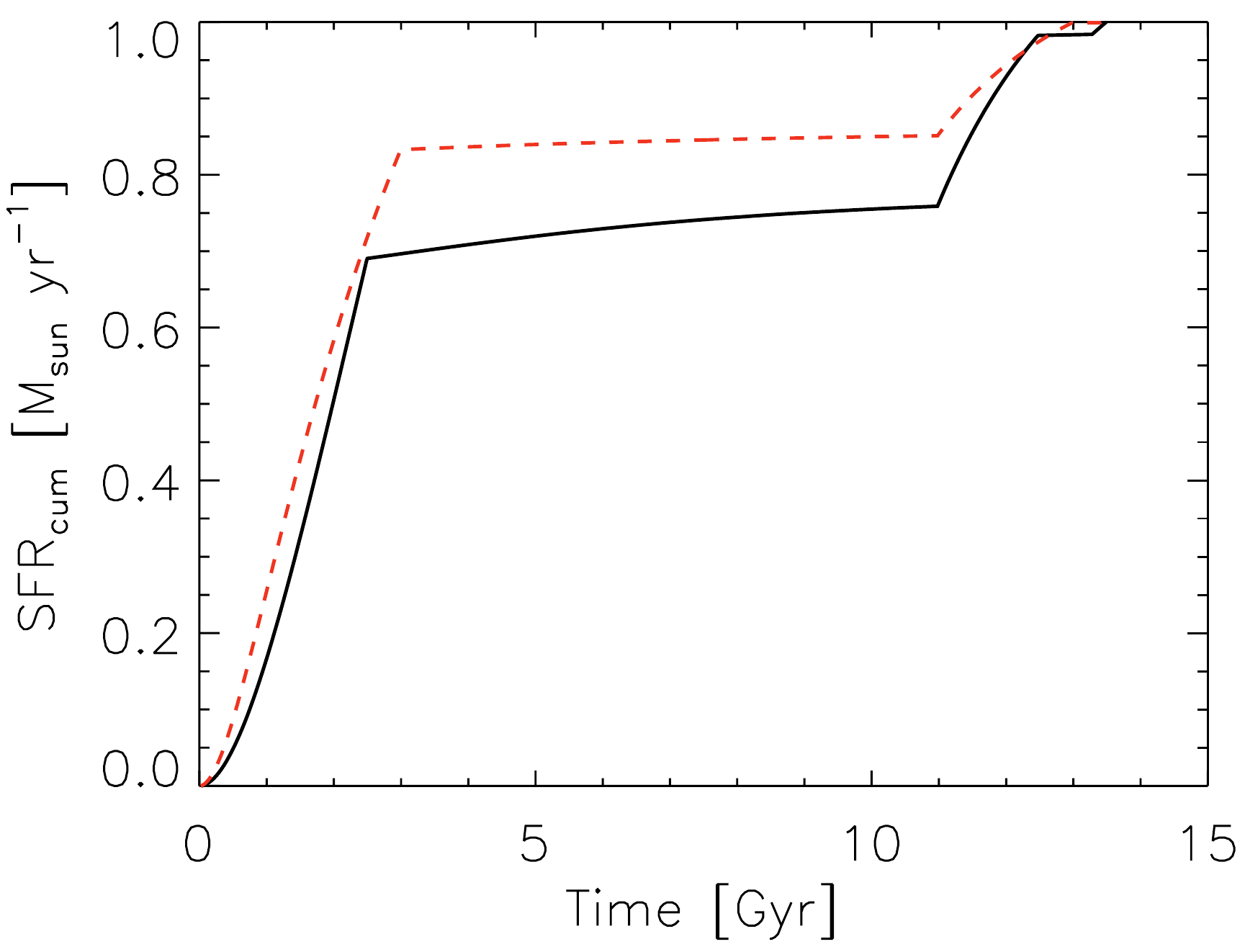}   
\caption{Left: the two SFHs for NGC\,185 used in \citet{2012MNRAS.419.3159M} (left panel) and their corresponding cumulative SFH (right panel).}
             \label{SFH_NGC185}
\end{figure*}

Based on prescriptions of \citet{1976ApJ...204..365F}, \citet{1998ApJ...507..726S} and \citet{1998ApJ...499..209W} estimated the gas mass returned to the ISM by planetary nebulae in the three dwarf spheroidal galaxies NGC\,147 (6-11$\times$10$^{5}$\,M$_{\odot}$), NGC\,185 (8-17$\times$10$^{5}$\,M$_{\odot}$) and NGC\,205 (23 $\times$10$^{5}$\,M$_{\odot}$), which are similar to or in excess of the current gas content in those galaxies. As a proof of concept, we redo these calculations for the dwarf spheroidal NGC\,185\footnote{We refrain from redoing the calculations for NGC\,147 and NGC\,205 due to the lack of sufficient constraints on their recent star formation histories.} based on a simple chemical evolution model with a closed-box approximation to account for the gas and dust mass returned by planetary nebulae and supernovae. In this simple model, we calculate the gas and dust mass that has been returned by the intermediate age population (2-3\,Gyr) based on the best fitting star formation histories (SFH) presented by \citet{2012MNRAS.419.3159M} which were optimised to fit the abundance ratios of PNe, the age-metallicity relation and the total galaxy mass at the present day. Figure \ref{SFH_NGC185} shows their two best fitting SFHs (left panel) and the cumulative SFHs (right panel). The latter corresponds well to the cumulative SFH presented by \citet{2015ApJ...811..114G} that was derived from colour-magnitude diagrams based on deep $V$ and $I$ band \textit{Hubble Space Telescope} ACS observations. Also the average current SFR$\sim$2.9-3.8$\times$10$^{-3}$ M$_{\odot}$ yr$^{-1}$ measured over a time period of 1\,Gyr, is consistent with the SFR derived by \citet{1999AJ....118.2229M} (SFR$\sim$6.6$\times$10$^{-4}$\,M$_{\odot}$ yr$^{-1}$) considering that the latter value only accounts for star formation that took place in the central regions of NGC\,185.
 
During the simulation, we track the gas consumption, dust production and return of gaseous material to the ISM based on these SFHs at individual time steps of 10\,Myr. At every time step, the contribution from stars with lifetimes $\tau_{m}$ = t - t$_{0}$ (with $t_{\text{0}}$ the age at which the galaxy was born) is taken into account. The stellar lifetimes for stars of a different mass and metallicity are calculated based on the parametrisation of \citet{1996A&A...315..105R}. The dust yields are taken from \citet{2008A&A...479..453Z} and \citet{2007MNRAS.378..973B} for intermediate (0\,M$_{\odot}$$<$M$<$8\,M$_{\odot}$) and high mass (12\,M$_{\odot}$$<$M$<$40\,M$_{\odot}$) stars, respectively. The gas yields from \citet{1997A&AS..123..305V} and \citet{1995ApJS..101..181W} are used for intermediate mass and massive stars, respectively. We interpolate between the intermediate and high mass estimates to derive dust and metal yields for stars with masses 8\,M$_{\odot}$$<$M$<$12\,M$_{\odot}$. We neglect stars more massive than 40\,M$_{\odot}$ since they will collapse to form black holes at the end of their lives and have a negligible contribution to the enrichment of the ISM. For our calculations, we assume a \citet{1955ApJ...121..161S} IMF with a slope of -2.35 within a mass range from 0.1\,M$_{\odot}$ to 100\,M$_{\odot}$.

Running the simulation with an initial gas mass of $M_{\text{gas}}$(t=0)=10$^{7.5}$\,M$_{\odot}$ and metal abundance [O/H](t=0)$\sim$10$^{-4}$ (consistent with the initial conditions used by \citealt{2012MNRAS.419.3159M}), we derive a gas mass (3-6$\times$10$^{5}$\,M$_{\odot}$) returned to the ISM since the second burst of star formation initiated $\sim$3.5\,Gyr ago until the current epoch. The total gas reservoir at the current epoch is estimated to be 1.1$\times$10$^{6}$\,M$_{\odot}$. With an observed gas mass (1.9-5.5$\times$10$^{5}$\,M$_{\odot}$, see Section \ref{Totalgas.sec}) that is two to five times smaller than predicted by our simple closed box model (and up to 3 times smaller than the gas mass returned to the ISM during the last two SF episodes), we argue that the closed box approximation does not fit the observational constraints. Similarly, \citet{2012MNRAS.423.2359D} showed that the current gas mass reservoir in NGC\,205 is too low compared to predictions of the gas mass returned by planetary nebulae. The non-detection of a gaseous reservoir in NGC\,147 furthermore seems unlikely given the population of evolved stars in this galaxy \citep{2005AJ....130.2087D}. Based on a comparison of the observed dust and gas reservoirs in NGC\,185 with a simple closed box model, we argue the low gas-to-dust mass ratio and gas deficiency result from gas removal processes. This gas removal could be induced by internal mechanisms (e.g., supernova explosions, stellar winds) and/or tidal interactions (see Section \ref{Tidal.sec}).

In \citet{2016MNRAS.459.3900D} we showed that the observed dust content is higher compared to the dust mass produced by AGB stars and supernovae during the last 100\,Myr (which is the estimated dust survival time) in NGC\,185. Based on the closed box model presented in this paper, we predict the production of up to 400\,M$_{\odot}$ during the last 100\,Myr. To account for the observed dust mass in NGC\,185 (5.1$\times$10$^{3}$\,M$_{\odot}$), we would require an efficient dust production during the last 1.6\,Gyr without any grain destruction. The latter dust survival time is significantly higher than the estimated dust lifetime in NGC\,185 (50-100\,Myr). In \citet{2016MNRAS.459.3900D}, we had argued that grain growth in the dense ISM phases could be an additional source of dust production, but the mechanisms that would enable the accretion of material onto grain surfaces in the ISM are not well understood \citep{2016arXiv160607214F}. Other than longer dust survival times, the metal production in current nucleosynthesis models, and the dust yields of AGB stars and supernovae in dust nucleation models might be underestimated.

\subsection{Gas removal}
\label{Tidal.sec}

Gas removal can result from internal mechanisms (e.g., supernova explosions, stellar winds), or external influences (e.g., hydrodynamical or gravitational interactions). Based on analytic/numerical models for dark matter halos with the inclusion of stellar feedback \citep{2000MNRAS.313..291F}, a total blow-away of the entire gaseous medium is only possible for dark matter halos of $M_{\text{g}}$ $\sim$ 5 $\times$ 10$^{6}$ M$_{\odot}$ \citep{2000MNRAS.313..291F}. With galaxy masses of $M_{\text{g}}$ $\sim$ 7.2 $\times$ 10$^{8}$ M$_{\odot}$ and $M_{\text{g}}$ $\sim$ 5.6 $\times$ 10$^{8}$ M$_{\odot}$ \citep{2010ApJ...711..361G} for NGC\,185 and NGC\,147, the dwarf spheroidal galaxies might lose some (but not all) gas to the intergalactic medium. The latter scenario is consistent with chemical evolution models which require efficient galactic winds to reproduce observed gas masses and abundances for the dSph galaxy population (e.g., \citealt{2004MNRAS.351.1338L,2010A&A...512A..85L,2012MNRAS.419.3159M}). Some of the metal-enriched gas expelled by galactic winds is assumed to rain back down on the galaxy disk according to a ``galactic fountain" mechanism that is able to flatten the metallicity gradient in those dSphs \citep{1994ApJ...431..598D,2000MNRAS.313..291F,2002A&A...394L..15B}. 

While this ``galactic fountain" effect might work for heavier galaxies, we argue that any gas expelled from these Andromeda dSph dwarfs will easily escape from the galaxy if heated to sufficiently high temperatures. Using the total galaxy masses (including baryonic and dark matter) for NGC\,147, NGC\,185 and NGC\,205 from \citet{2006MNRAS.369.1321D} within 2 effective radii ($R_{\text{eff}}$), we derive escape velocities\footnote{The escape velocity is calculated from $v_{\text{esc}}$~=~$\sqrt{2GM/R}$ using the total galaxy mass, M, the distance to the centre of mass, $R$=2*$R_{\text{eff}}$, and the gravitational constant, G.} that range from 57 to 91 km s$^{-1}$ at a radius of 2*$R_{\text{eff}}$. By heating the gas to sufficiently high temperatures, the thermal gas velocity of a gas with a temperature of $T_{\text{kin}}$~=~10$^6$\,K ($v_{\text{th}}$ = 90 km s$^{-1})$\footnote{The thermal gas velocity for a gas with kinetic temperature $T_{\text{kin}}$~=~10$^6$\,K is calculated from $v_{\text{th}}$ = $\sqrt{k_{\text{B}}*T/m}$ using the Boltzmann constant, $k_{\text{B}}$, the mass of a hydrogen atom, m, and the gas temperature $T_{\text{kin}}$~=~10$^6$\,K.} would be sufficient for the gas no longer to be gravitationally bound to the galaxy. This simple calculation shows that if the hot gas is blown out by supernova feedback and/or stellar winds to large radii, it might be able to escape from the galaxy if heated to sufficiently high temperatures. The typical hot X-ray halo of gas that provides more massive galaxies with fresh gas supplies for star formation does not seem to be present in those lower metallicity dwarfs which is observationally supported by the non-detections of X-ray emission in these dwarfs (see Section \ref{X-ray.sec}). The presence of dust and the metal enrichment (0.2-0.3 dex) over the last $\sim$ 8Gyr \citep{2012MNRAS.419..854G} in the central regions of NGC\,185 is consistent with the absence of ``galactic fountains" which would distribute the metals throughout the galaxy's disk.

But this observed central concentration of metals is also compatible with a scenario of external influences that mostly remove the metal-poor H~{\sc{i}} gas from the outer galaxy parts \citep{1990ApJ...357..367V}. Tidal interactions with other satellite galaxies or Andromeda can also potentially remove the gas from the outer galaxy regions in NGC\,185 and entirely strip the gas from NGC\,147. Recent observations from the Pan-Andromeda Archaeological Survey (PAndAS) reveal isophotal twisting and the emergence of extended tidal tails in NGC\,147, but do not show any evidence for tidal effects on the stellar light profiles of NGC\,185 \citep{2014MNRAS.445.3862C}. The asymmetric H~{\sc{i}} distribution, combined with its small extent up to only 1/4th of its Holmberg radius \citep{1997ApJ...476..127Y} might be an indication for tidal effects having played an important role in the evolution of NGC\,185 in the past. Being located at an angular distance of 12 degrees from M\,31, NGC\,185 and NGC\,147 are currently beyond the tidal influence radius of Andromeda\footnote{For NGC\,185 and NGC\,147, the tidal radius has been calculated to be between 10-12 kpc or, thus, beyond 25 effective radii \citep{2010ApJ...711..361G}.}. Given their small angular separation ($\sim$1$^{\circ}$), the two galaxies have been argued to form a gravitationally bound pair \citep{1998AJ....116.1688V,2010ApJ...711..361G}. Based on their carbon star populations \citep{2004A&A...417..479B}, the different ages of the dominant old stellar populations (suggesting their separate infall into the group system, \citealt{2015ApJ...811..114G}), timing arguments \citep{2013MNRAS.430..971W}, and conditions for their gravitational bound \citep{2014MNRAS.440.1225E}, there is however no reason to assume that the two dwarf spheroidals form a close pair. \citet{2013MNRAS.430..971W} rather suggest a close connection between NGC\,185 and Cass II. The possible bound with the dwarf spheroidal galaxy Cass II might have had an influence on the recent star formation in NGC\,185. Earlier interactions with Andromeda (or even the Milky Way, \citealt{2012MNRAS.426.1808T}) could be responsible for significant gas removal in the past. It is, for the moment, unclear whether tidal interactions with M\,31 or mutual encounters between the dwarf satellites (NGC\,147, NGC\,185, Cass II) have caused the gas stripping in the two galaxies. In NGC\,205, there is observational evidence for a past tidal interaction with the Andromeda galaxy that could have removed part of the gas content of NGC\,205 (see \citealt{2012MNRAS.423.2359D} for more details). Better knowledge on the orbits of the three dwarf spheroidal galaxies is required to model the past interactions with companions in the Andromeda group.

\subsection{Galaxy evolution}
\label{Miss.sec}

Although the three dwarf spheroidal galaxies probably share a similar evolutionary history (driven by galactic winds and/or tidal interactions), their different ISM conditions (i.e., central ongoing star-formation in NGC\,185 and NGC\,205, and the lack of any detectable ISM material in NGC\,147) shows that similar mechanisms can result in a variety of morphological outcomes (e.g., \citealt{2013MNRAS.428.2980R}) depending on the efficiency of galactic winds and the orbit of the galaxy (e.g., \citealt{2011ApJ...726...98K}). 

Chemical evolution models seem to require high galactic wind efficiencies to explain the build up of gas and metals in these galaxies (e.g., \citealt{2004MNRAS.351.1338L,2010A&A...512A..85L,2012MNRAS.419.3159M}), while tidal stirring \citep{2001ApJ...559..754M} and galaxy threshing \citep{2001ApJ...552L.105B} have been put forward as the most important mechanisms for the formation of dwarf spheroidals and ultracompact dwarfs in low-density group environments based on galaxy simulations. Also observational evidence of tidal influence for galaxies residing in group environments (e.g., \citealt{2014ApJ...796L..14P}) supports these theoretical simulations. 

Given the wide range of resulting end products and the continuous influence of environmental effects on most galaxies residing in cluster and group environments, it has hard to constrain the progenitor galaxies of these dwarf spheroidal galaxies in the Local Group \citep{2013MNRAS.432.1162L}. Rather than the transformation of dwarf irregular into dwarf elliptical galaxies, the present-day dwarf galaxy population might originate from the same common progenitor population that experienced a different evolution due to differences in dark matter content, stellar mass and/or environments (e.g., \citealt{2000MNRAS.313..291F,2009ARA&A..47..371T,2012MNRAS.420.1714S}).
The only way to properly constrain the evolutionary history of the population of dwarf spheroidal galaxies that is present-day observed in group and cluster environments, is through a combination of observations probing their stellar populations, star formation history, chemical enrichment, kinematic properties (e.g., \citealt{2009ARA&A..47..371T}) and orbital parameters (e.g., \citealt{2008ApJ...683..722H,2013MNRAS.430..971W}).

\section{Conclusions}
\label{Conclusions.sec}
We make an inventory of the gas content in three low-metallicity dwarf spheroidal galaxies of the Local group (NGC\,147, NGC\,185 and NGC\,205) based on an extensive set of ancillary observations. We present new Nobeyama CO(1-0) observations that cover the previously unexplored regions in the south of NGC\,205, and we use \textit{Herschel} SPIRE FTS [C~{\sc{i}}] observations to limit the fraction of CO-dark gas in NGC\,205. Based on \textit{Herschel} observations of the far-infrared fine-structure lines [C~{\sc{ii}}] and [O~{\sc{i}}] towards the central regions NGC\,185, we analyse the typical conditions of the ISM in dSphs.  

We compute total gas masses of $M_{\text{g}}$ = 1.9-5.5$\times$10$^{5}$\,M$_{\odot}$ (NGC\,185) and $M_{\text{g}}$ = 8.6-25.0$\times$10$^{5}$\,M$_{\odot}$ (NGC\,205) within the limits of uncertainty on the $X_{\text{CO}}$ factors, by combining the mass reservoirs of atomic, cold and warm molecular, ionised and hot X-ray gas. Non-detections result in an upper gas mass limit of $M_{\text{g}}$ $\leq$ 0.3-2.2$\times$10$^{5}$ M$_{\odot}$ for NGC\,147. Our new NRO 45m CO(1-0) map of the southern regions in NGC\,205 shows that most of the molecular gas is distributed towards the north and centre. The non-detections of the [C~{\sc{i}}] 1-0 and 2-1 line transitions in the SPIRE FTS spectra implies that the CO-dark gas fraction is negligible in NGC\,205 compared to the molecular gas mass traced by CO, which is also consistent with the lower [C~{\sc{ii}}]/CO ratios (2-4$\times$10$^{3}$) in dSphs compared to low-metallicity star-forming dwarf galaxies with [C~{\sc{ii}}]/CO ratios of a few times 10$^{4}$. 

Photo-dissociation models suggest a soft radiation field (G$_{\text{0}}$$\sim$1-30) and moderate hydrogen gas density (n$_{\text{H}}$$\sim$10$^{3.75}$-10$^{4.25}$cm$^{-3}$) to explain the observed [C~{\sc{ii}}], [O~{\sc{i}}] and total-IR emission in NGC\,185. The detection of several high excitation lines implies that also a dense PDR phase with small filling factor is present, or alternatively requires shocks to excite the lines. The high [C~{\sc{ii}}]/TIR$\sim$1.5$\%$ and [C~{\sc{ii}}]+[O~{\sc{i}}]/TIR$\sim$2$\%$ ratios indicate that the photoelectric efficiency is high, which might be explained by a high PAH abundance and/or low level of grain charging in NGC\,185. The star formation rate densities and current gas reservoirs in NGC\,185 and NGC\,205 places these galaxies above the main sequence of star forming galaxies. The short molecular gas depletion time scales imply that fuel for star formation will run out in less than a few 100\,Myr in these dSphs. 

We derive global gas-to-dust mass ratios of GDR$\sim$37-107 and GDR$\sim$48-139 which are at the low end of the average Milky Way ratio of GDR$\sim$130 and significantly lower compared to the expected ratios of GDR$\sim$370 and GDR$\sim$520 for the metal abundances in NGC\,185 (0.36\,Z$_{\odot}$) and NGC\,205 (0.25\,Z$_{\odot}$), respectively. Based on a simple closed box model, we confirm that these dSphs are gas deficient and that the dust has a longer dust survival time ($\sim$1.6\,Gyr) in these galaxies which can also explain their anomalous GDR. We conclude that part of the gas content has been removed from the dSph satellites in the recent past. We believe that efficient galactic winds (combined with the heating of gas to sufficiently high temperatures in order for it to escape from the galaxy) and/or environmental interactions with neighbouring galaxies are responsible for the gas removal from NGC\,147, NGC\,185 and NGC\,205.

\section*{Acknowledgments}
The authors would like to thank Marla Geha and Martha Boyer for interesting discussions that have helped to improve this paper.
We would like to thank Denise Gon{\c c}alves and Laura Magrini for kindly sharing their H$\alpha$ data presented in \citet{2012MNRAS.419..854G}.
PACS has been developed by a consortium of institutes
led by MPE (Germany) and including UVIE
(Austria); KU Leuven, CSL, IMEC (Belgium);
CEA, LAM (France); MPIA (Germany); INAFIFSI/
OAA/OAP/OAT, LENS, SISSA (Italy);
IAC (Spain). This development has been supported
by the funding agencies BMVIT (Austria),
ESA-PRODEX (Belgium), CEA/CNES (France),
DLR (Germany), ASI/INAF (Italy), and CICYT/
MCYT (Spain). SPIRE has been developed
by a consortium of institutes led by Cardiff
University (UK) and including Univ. Lethbridge
(Canada); NAOC (China); CEA, LAM
(France); IFSI, Univ. Padua (Italy); IAC (Spain);
Stockholm Observatory (Sweden); Imperial College
London, RAL, UCL-MSSL, UKATC, Univ.
Sussex (UK); and Caltech, JPL, NHSC, Univ.
Colorado (USA). This development has been
supported by national funding agencies: CSA
(Canada); NAOC (China); CEA, CNES, CNRS
(France); ASI (Italy); MCINN (Spain); SNSB
(Sweden); STFC and UKSA (UK); and NASA
(USA). 
This research has made use of the NASA/IPAC Extragalactic Database (NED) which is operated by the Jet Propulsion Laboratory, California Institute of Technology, under contract with the National Aeronautics and Space Administration.




\bsp	
\label{lastpage}
\end{document}